\documentclass[preprint,11pt]{aastex}
\usepackage{graphicx}
\usepackage{natbib}
\shorttitle{The Search for Planetary Systems with SIM}
\shortauthors{Sozzetti et al.}

\begin{document}


\title{Narrow-Angle Astrometry with the Space Interferometry Mission: \\
       The Search for Extra-Solar Planets. \\
       II. Detection and Characterization of Planetary Systems}


\author{A. Sozzetti\altaffilmark{1,2,3}}
\altaffiltext{1}{University of Pittsburgh, Dept. of Physics \&
Astronomy, Pittsburgh, PA 15260, USA} \altaffiltext{2}{Smithsonian
Astrophysical Observatory, Harvard-Smithsonian Center for
Astrophysics, 60 Garden Street, Cambridge, MA 02138}
\altaffiltext{3}{Osservatorio Astronomico di Torino, 10025 Pino
Torinese, Italy}\email{alex@phyast.pitt.edu}

\author{S. Casertano\altaffilmark{4} and R. A. Brown\altaffilmark{4}}
\altaffiltext{4}{Space Telescope Science Institute, Baltimore, MD
21218, USA}\email{stefano@stsci.edu; rbrown@stsci.edu}

\and

\author{M. G. Lattanzi\altaffilmark{3}}
\email{lattanzi@to.astro.it}



\begin{abstract}

We utilize $a)$ detailed end-to-end numerical simulations of
sample narrow-angle astrometric observing campaigns with the Space
Interferometry Mission (SIM) and the subsequent data analysis
process, and $b)$ the set of extra-solar planetary systems
discovered so far by radial velocity surveys as templates to
provide meaningful estimates of the limiting capabilities of SIM
for the detection and measurement of multiple-planet systems
around solar-type stars in the solar neighborhood.

We employ standard $\chi^2$ statistics, periodogram and Fourier
analysis to evaluate SIM's ability to detect multiple planetary
signatures: the probability of detecting additional companions
is essentially unchanged with respect to the single-planet
configurations, but after fitting and subtraction of orbits with astrometric
signal-to-noise ratio $\alpha/\sigma_d\rightarrow 1$ the false
detection rates can be enhanced by up to a factor 2; the
periodogram approach results in robust multiple-planet detection 
for systems with periods shorter than the SIM mission length, even at low
values of $\alpha/\sigma_d$, while the least squares technique combined
with Fourier series expansions is arguably preferable in the
long-period regime. We explore the three-dimensional parameter
space defined by astrometric signature, orbital period, and
eccentricity to derive general conclusions on the capability of
SIM to accurately measure the full set of orbital parameters and
masses for a variety of configurations of planetary systems: the
accuracy on multiple-planet orbit reconstruction and mass
determination suffers a typical degradation of 30-40\% with respect to
single-planet solutions; mass and orbital inclination can be
measured to better than 10\% for periods as short as 0.1 yr, and
for $\alpha/\sigma_d$ as low as $\sim 5$, while $\alpha/\sigma_d\simeq 100$ 
is required in order to measure with similar accuracy systems 
harboring objects
with periods as long as three times the mission duration. We gauge
the potential of SIM for meaningful coplanarity measurements via
determination of the true geometry of multiple-planet orbits: for
systems with all components producing $\alpha/\sigma_d\simeq 10$
or greater, quasi-coplanarity can be reliably established 
with uncertainties of a few degrees, for 
periods in the range $0.1\leq T\leq 15$ yr; in systems where at least one
component has $\alpha/\sigma_d\rightarrow 1$, coplanarity 
measurements are compromised, with typical uncertainties on 
the mutual inclinations of order of $30^\circ-40^\circ$. 
We quantify the
improvement derived in full-orbit reconstruction and planet mass
determination by constraining the multiple-planet orbital fits to
SIM observations with the nominal orbital elements obtained from
the radial velocity measurements: the uncertainties on orbital
elements and masses can be reduced by up to an order of
magnitude, especially for long-period orbits in face-on
configurations, and for low amplitude orbits seen edge-on.

Our findings are illustrative of the importance of the contribution
SIM will make, complementing other on-going and planned spectroscopic,
astrometric, and photometric surveys, in order to fulfill the
expectations for ground-breaking science in the fields of formation and
evolution of planetary systems during the next decade.

\end{abstract}


\keywords{astrometry -- planetary systems -- instrumentation:
interferometers -- methods: data analysis -- methods: numerical}

\section{Introduction}

After more than thirteen years of precise radial velocity
measurements, today there is clear evidence that several nearby
solar-type stars harbor candidate planetary systems, composed of
two or more planets~\citep{butler99,marcy01b,mayor01,jones02,fischer02,
marcy02b,fischer03,butler03}, or a planet
and a probable brown dwarf~\citep{marcy01a,udry02} 

The answers provided by these discoveries have presented even
more challenging puzzles to theoretical models aimed at describing
the formation and evolution of planetary systems. The variety of
orbital arrangements in multiple-planet systems found by spectroscopy
calls into question both the origin and early dynamical evolution of
such systems, in terms of formation mechanisms and orbital migration
scenarios, as well as their long-term dynamical evolution and stability.
Early attempts have been made to connect and relate the early
stages of their formation to their presently observed orbital properties.
For example, the two planets orbiting the M4V star Gliese 876 appear to be
locked in a 2:1 resonance, and the existence of such commensurabilities
may be considered important evidence of possible inward migration for pairs
of planets due to mutual interactions as well as tidal interaction with the
protoplanetary disk during and shortly 
post-formation~\citep{snell01,nelson02}. 
Similarly, the three-planet system around the F8V star $\upsilon$ 
Andromed\ae\, ($\upsilon$ And) appears to be stable over evolution 
timescales of order of the star's lifetime, again suggesting the possibility
that the system's orbits were shaped by the interaction of protoplanets
with the protoplanetary disk, in the first few million years of $\upsilon$
And's existence~\citep{arty01}. Furthermore, in many of the observed
multiple-planet systems the eccentricities are much larger than
those of the planets of our own solar system. Again, this may be
explained with significant eccentricity evolution due to mutual interaction
between planets as well as with the protoplanetary disk during the
epoch of disk depletion~\citep{chiang02,nagasawa03}.
However, direct extrapolations from the early stages of their formation to
the presently observed configurations of extra-solar planetary systems are
still somewhat speculative. Today, in fact, there is still a lack of
observational support to the theoretical models describing the early epochs
of the formation of planetary systems, either because of the objective
difficulty to identify good targets among very young stellar objects
(often completely obscured by circumstellar material) or due to
insufficient sensitivity of the present generation
of ground-based and space-borne instrumentation. Thus, theorists have
focused primarily on the issues of long-term
stability and orbital evolution of extra-solar planetary systems,
independent on the details of their formation.

Since Newton's discovery of the law of universal gravitation, the problem
of predicting the orbits of the planets in our own solar system has been 
tackled by many giants of mathematics (Euler, Laplace, Jacobi, Lagrange, 
Gauss, Poincar\'e). By the end of the nineteenth century, 
it had been demonstrated
that it is not possible to find an exact analytic solution describing the
motion of $N$ bodies, for $N\ge 3$, and the theory of perturbations in
celestial mechanics was developed and extensively applied to many
interesting cases, always within the boundaries of our solar system. During
the two last decades, with the advent and exponential progress of
computer technology, large-scale
computations that follow the full evolution of the nine planets for a
significant fraction of the lifetime of the Sun became possible. This
has allowed the establishment of, 
for example, the long-term chaotic behavior of the
orbits of planets in the solar system~\citep{sussman92}. 
In fact, the typical (Lyapunov) timescale for exponential divergence of two 
orbits in the solar system starting with slightly different initial 
conditions is only several million years, and soon the issue of stability 
only becomes a statistical question. The planetary orbits in the solar 
system are on the other hand stable, in the sense that the probability of
ejections or close encounters during a time frame comparable to the
main-sequence lifetime of the Sun ($\sim 12$ Gyr) is extremely low.

However, until the first multiple-planet system was announced
~\citep{butler99}, the problem of planetary orbits dynamics had
in recent years been almost relegated to the level of an academic
pursuit. Now that a variety of additional systems, with a wide
range of different orbital arrangements, has been found, the field
of planetary system dynamics has received revived interest. The
long-term dynamical evolution of the 11 extra-solar planetary
systems known to date has been extensively investigated by
theorists, utilizing both direct numerical integrations and
approximate analytical approaches for the description of the
N-body mutual interactions. As opposed to our own solar system, or
the three-planet system discovered by Wolszczan \& Frail~\cite{wols92}
around the pulsar PSR 1257+12, both appearing dynamically stable
over long timescales~\citep{laskar94,gladman93}, the dynamical
evolution and long-term stability issues for the majority of the
multiple-planet systems discovered by radial velocity are highly 
uncertain. As a general result, studies
specifically targeted to gauge the general dynamical behavior of 
extra-solar planetary systems such as $\upsilon$ And
~\citep[Chiang \& Murray 2002]{laugh99,rivera00,
stepinski00,jiang01,barnes01,chiang01,rivera01a}, 
55 Cancri~\citep[55 Cnc~hereafter.][]{novak02,ji03}, 47
Urs\ae\, Majoris~\citep[47 UMa~hereafter.][]{godz02,laugh02}, Gliese 876
~\citep[2002b]{kino01,laugh01,rivera01b,ji02,godzetal02,lee02a}, HD
82943, HD 37124, HD 12661, HD 38529, and HD 160691 (God\'zdziewski 
\& Maciejewski 2001; Kiseleva-Eggleton et al. 2002; God\'zdziewski 
\& Maciejewski 2003; God\'zdziewski 2003a, 2003b), come to the same 
conclusion: the stability of the systems can be greatly affected by small
variations of the instantaneous orbital elements of the planetary
orbits obtained from the radial velocity data and utilized as
initial conditions for the numerical integrations. In particular,
for the systems not to be destabilized and disrupted on very short
timescales (as short as 10$^3$-10$^5$ years), constraints must be
placed on the maximum allowed values for the masses and on the
range of allowed relative inclinations of the orbits. Often
analytical calculations are performed with the coplanarity of the
orbits as a working assumption. Early attempts have also been made
to verify the possibility of regions of dynamical stability inside
the parent stars' Habitable Zones~\citep{kasting93}, where Earth-size 
planets may be found (Jones et al. 2001; Noble et al. 2002; Jones \& 
Sleep 2002a, 2002b; Rivera \& Haghighipour 2002; Cuntz et al. 2003; 
Menou \& Tabachnik 2003). However, the existence of such
regions does not directly imply that rocky planets may have
actually formed, in these systems, at such privileged distances,
in the first place~\citep{thebault02}. 

The results obtained by many of the present studies on the
dynamical stability of extra-solar planetary systems contain
significant ambiguities. The impossibility to draw at present more
than general statistical conclusions on such issues arises partly
from the intrinsically chaotic nature of N-body systems, and
partly from a lack of knowledge of a few key parameters that
cannot be derived by the present observational datasets, but must
be {\it a priori} fixed throughout the analyses. As a matter of
fact, in general four quantities drive the effects of the mutual
perturbations in multiple-planet systems investigated by
theorists: the planetary eccentricities, the orbital periods (and
possible commensurabilities among them), the planet masses
(usually through their ratio), and the relative inclinations of
the orbital planes. The last two of these parameters cannot be
determined from the intrinsically one-dimensional information
extracted from stellar spectra. In fact, radial velocity
measurements do not determine either the inclination $i$ of the
orbital plane with respect to the plane of the sky (and this in
turn allows only for lower limits to be placed on the actual mass
of each companion to the observed star) or the position angle
$\Omega$ of the line of nodes in the plane of the sky (and thus
relative inclination angles remain unknown): without knowledge of
the full three-dimensional geometry of the systems and true mass
values, general conclusions on the architecture, orbital evolution
and long-term stability of the newly discovered planetary systems
remain questionable.

Attempts have been made to break the $\sin i$ degeneracy and
determine the true masses of planets in multiple systems utilizing
self-consistent dynamical fits~\citep{laugh01,rivera01b}. 
Such procedures significantly
improve orbital fits performed assuming independent Keplerian motions
in cases such as that of Gliese 876, where mutual perturbations are relevant
and the orbital elements of the planets undergo significant changes on
timescales comparable to the timespan of the observations. However,
these dynamical fits to the radial velocity data are not conclusive,
as assumptions
must still be made on the actual relative inclinations of the planets,
and broad ranges of $\sin i$ provide similarly good fits. Thus,
tightly constraining planetary masses will be difficult for several years
to come, if one has to rely only upon radial velocity measurements.
Furthermore, different orbital configurations giving similarly
good values of the reduced chi-square $\chi^2_\nu = \chi^2/\nu$
(where $\nu$ is the number of degrees of freedom) behave very differently
when integrated over long timescales, and the ambiguity on the actual
long-term stability of the systems cannot be removed.

In the latest years other techniques have reached the sensitivity
necessary to 
complement spectroscopy in extra-solar planet searches. In particular,
transit photometry has been successful in identifying the first gas
giant planet eclipsing its parent star~\citep{charbon00,henry00}, 
confirming the spectroscopic detection~\citep{mazeh00,queloz00}, 
and very recently in directly discovering 
the second transiting planet~\citep{konacki03}. The 
observations of planetary transits on the disk of HD 209458 allow
for a direct estimate of the planet's size and actual mass (as the
inclination of the orbital plane is known, and the $\sin i$ degeneracy
breaks down), thus, also thanks to the recent detection of sodium in
its atmosphere~\citep{charbon02}, its density can be inferred
and important insights on its composition can be obtained. However, transit
photometry detects only the small proportion of planets whose orbits happen
to line up almost exactly with the line of sight to the star.
We anticipate that future high-precision ground-based
~\citep{mariotti98,booth99,colavita99} as well as space-borne
~\citep{danner99,perryman01} astrometric observatories
will be among the most effective techniques 
to remove the present ambiguities in the dynamical
analysis of extra-solar planetary systems as well as provide valuable
data to probe models addressing the issues of the formation of planetary
systems and the actual nature of sub-stellar companions.
The intrinsically two-dimensional astrometric data provide the
means to directly measure the two missing parameters, i.e.
the inclination angle and the line of nodes of each planetary orbit.
Knowledge of the full viewing geometry of a system of planets allows
then to derive meaningful estimates for the true masses of each
orbiting object and the relative inclinations of the orbital planes.

In our previous work~\citep[S02~hereafter]{sozzetti02}, 
we addressed the issues of the detectability and measurability of
single planets around single stars with the Space Interferometry
Mission (SIM), with the instrument operated in narrow-angle
astrometric mode. We expressed our results as a function of both
SIM mission parameters and properties of the planet (mass, orbital
characteristics). In the continuation of our studies we have
utilized extensive simulations of SIM observations of the present
sample of extra-solar multiple-planet systems in order to draw
general conclusions on the ability of SIM to discover and measure
systems of planets, as well as quantify the instrument's
capability to determine the coplanarity of multiple-planet orbits.
Similarly, Sozzetti et al.~\citeyearpar{sozzetti01} 
have recently assessed the capabilities of ESA's 
Cornerstone Mission GAIA for the detection and measurement of
planetary systems utilizing the $\upsilon$ And system as a
template. Our study is different in that, addressing the entire
set of presently known multiple-planet systems, it extends the
analysis beyond the favorable cases (well-spaced, well-sampled
orbits, high ``astrometric'' signal-to-noise ratios) studied by
Sozzetti et al.~\citeyearpar{sozzetti01}, and is specifically 
tailored to SIM.

This second paper is organized as follows. In the second Section
we briefly describe the setup for the simulation of SIM sample
narrow-angle campaigns, the statistical tools implemented for
planet detection, and the algorithms for multiple orbital fits. In
the third Section we present results on multiple-planet systems
detection, multiple-planet orbit reconstruction, and coplanarity
analyses, and compare our results to those of previous studies. 
Finally, in the fourth Section we summarize our results and discuss 
our findings in the context of the present status of planet searches. 

\section{Simulation Setup, Detection and Orbital Fitting Methods}

The code for the reproduction of sample observing campaigns with
SIM operated in narrow-angle astrometric mode and the subsequent
analysis of the simulated dataset has been thoroughly described in
S02, where we defined detectability horizons and limits on distance for
accurate orbital parameters and mass determination in the case of
single planets orbiting single, nearby solar-type stars.
In this Section we summarize its main features and working assumptions,
and describe the
modifications made in order to address the issues of multiple-planet
system observations, detection, and measurement.

\subsection{SIM Narrow-Angle Astrometric Observing Scenario}\label{scenario}

The SIM instrument~\citep[for a thorough presentation of the mission 
concept see for example][]{danner99} executes pointed observations 
at arbitrary times and 
orientations of the interferometric baseline. When operating in the 
regime of narrow-angle astrometry, 
its fundamental one-dimensional measurement
is the {\it relative} optical path-length delay:
\begin{equation}\label{relative}
  \Delta d_{\star,n} =
  \mathbf{B}\cdot(\mathbf{S}_{\star}-\mathbf{S}_n)+
  \sigma_d
\end{equation}
which corresponds to the instantaneous angular distance between
the target and its $n$-th reference star, projected onto the baseline of
the science interferometer, while the two guide interferometers
accurately monitor changes in the satellite's attitude by acquiring
fringes of two bright guide stars. In the above formula
$\mathbf{B} = B\mathbf{u}_\mathrm{b}$ is the
interferometer baseline vector of length $B$, $\mathbf{S}_{\star}$  and
$\mathbf{S}_n$ are the unit vectors to the target star and its
$n$-th reference object, while $\sigma_d$
is the single-measurement accuracy on each relative delay measurement.

Throughout all our simulations we have adopted the template 
observing scenario outlined in S02: within a domain of $1^\circ$ 
in diameter around each target we place $N_r = 3$ reference
stars, and execute $N_o = 24$ full two-dimensional observations
(corresponding to a total of $N_m = 144$ relative delay
measurements) randomly, uniformly distributed over the nominal $L
=5$ years mission lifetime, each composed of two one-dimensional
standard visits made with orthogonal orientations of the
interferometer baseline vector. Since in our study we have
utilized the present set of extra-solar planetary systems,
orbiting very bright ($V\leq 10$ mag), nearby, solar-type stars
(spectral types F-G-K, except for Gliese 876), we have utilized a
structure of the standard visit which applies to bright target and
reference stars ($V\leq 11$ mag), and consequently set a
single-measurement error $\sigma_d = 2$ $\mu$as
throughout all our analysis (see S02 for details on the adopted
error model). This value for $\sigma_d$ is in line with the
presently envisaged SIM performance in its new shared-baseline
configuration. \footnote{The results presented and discussed in
the following Sections can easily be rescaled to different timing 
and numbers of full-observations and reference stars, with the simple 
scaling laws already derived in S02}

The basic assumptions made in our previous work with regard to the SIM
instrument and the systems to be investigated are unchanged in this
study: $1)$ we assume perfect knowledge of the error model and attitude of
the spacecraft; $2)$ the objects composing the local frame of reference,
with respect to which observations of a given scientific target are made,
are assumed astrometrically clean, i.e. sources of astrometric noise
intrinsic to the source, such as flares, rotating spots, or
marginally unstable circumstellar disks, have
not been taken into consideration, nor have we discussed the possibility
of binaries among either targets or reference stars; $3)$ effects on
the targets' displacement due to perspective acceleration, changing parallax,
or higher-order contributions from relativistic aberration and light
deflection from the major solar system bodies have not been included.

We have modified the description of the target motion on the celestial
sphere to allow for the presence of multiple planets.
The stellar motion is described in terms of the 5 basic
astrometric parameters and of the gravitational perturbations
produced by the orbiting planets. The difference between
the position vector to the target $\mathrm{S}_{\star}$ evaluated
at time $t$ and the same quantity $\mathrm{S}^\prime_{\star}$
measured at time $t^\prime$ can then be expressed as a sum
of small perturbative terms:
\begin{equation}\label{model}
\mathrm{S}^\prime_{\star}-\mathrm{S}_{\star}={\rm d}{\bf
S}_{\star} = {\rm d}{\bf S}_{\mu}+{\rm d}{\bf S}_{\pi}+
\sum_{i=1}^{n_p}{\rm d}{\bf
S}_{K}^i,
\end{equation}
where $n_p$ is the number of planets orbiting the target. The gravitational
effects produced by the planets are assumed to be linear
(i.e., as the sum of independent Keplerian motions). Such
simplification (no mutual gravitational interactions between planets) has
in general little impact on the significance of our analysis.
In fact, assuming the orbits are coplanar, 
over the time-scale of SIM observations (5 years), variations
of the orbital parameters due to secular as well as resonant perturbative
terms can be confidently considered negligible. This assumption
may not be correct only in the case of Gliese 876, where the
two gas giant planets orbiting the star are locked in a 2:1 resonance, and
strong mutual perturbations are already evident in the data over the
timescale of the radial velocity observations ($\sim 13$ years). We plan
to study in detail in a future work the effects induced by mutual
gravitational perturbations between planets on the quality of
multiple-planet orbital fits that do and do not contain analytic
descriptions of the resonant and/or secular perturbative nature of the
systems.

\subsection{Statistical Tools for Planet Detection}\label{stattools}

We adopt two different procedures for investigating the efficiency of SIM
in detecting multiple-planet systems.

First of all, following the approach in S02, we employ a
standard $\chi^2$ test with confidence level set to 95\% applied to
the observation residuals. The test is performed initially against the
null hypothesis that there is no planet, then after removal of the
first planetary signature it is applied in succession until no further
significant deviations (at the 95\% confidence level) from the fitted
model containing $n_p$ planets appear in the residuals.

Secondly, we have upgraded our code to include a basic periodogram analysis
in order to search for hidden periodicities in the astrometric measurements.
To this aim, we have utilized the standard algorithm for the
evaluation of the Lomb-Scargle normalized periodogram for the spectral
analysis of unevenly sampled data~\citep{press92}, 
specifically its `fast' version by Press \& Rybycki~\cite{press89}.
The traditional Lomb-Scargle formula is equivalent to 
performing an unweighted linear
least-squares fit of a sinusoid to the data, after subtraction of the
mean. For a time series of relative delays $\Delta d(t_i)$, where $i$
takes on integer values up to the total number of measurements $N_m$, the
normalized periodogram as a function of the test angular 
frequency $\vartheta$ is defined as:
\begin{equation}
z(\vartheta) = \frac{1}{2\sigma_d^2}\left\{
\frac{\left[\sum_{i=1}^{N_m}(\Delta d(t_i)-\overline{\Delta d})
\cos\vartheta(t_i-t_c)\right]^2}{\sum_{i=1}^{N_m}\cos^2\vartheta(t_i-t_c)}+
\frac{\left[\sum_{i=1}^{N_m}(\Delta d(t_i)-\overline{\Delta d})
\sin\vartheta(t_i-t_c)\right]^2}{\sum_{i=1}^{N_m}\sin^2\vartheta(t_i-t_c)}
\right\}
\end{equation}
Here, $\overline{\Delta d}$ is the mean of the data, while $t_c$ is given by
\begin{equation}
t_c = \frac{1}{2\vartheta}
      \arctan\left(\frac{\sum_{i=1}^{N_m}\sin 2\vartheta t_i}
{\sum_{i=1}^{N_m}\cos 2\vartheta t_i}\right)
\end{equation}
The normalization factor is by default taken as the variance
of the sample. Because of the exploratory nature of this investigation,
we assign equal weight to all observations and set $\sigma_d = 2$ $\mu$as,
the value of the single-measurement error on relative delays as defined
in the previous Section.
At the moment we do not discuss the merit of other possible choices for the
normalization of the periodogram~\citep{gilli87,czerny96}.

Observation residuals are inspected successively after the single-star fit,
and after removal of each planetary signature, to establish the existence
or absence of further periodicities. Also in this case, the significance of
the detection or non detection of a periodic signal in the residuals,
i.e. the significance of a particular peak in the power spectrum
$z(\vartheta)$, must be assessed. For the above choice of the normalization
factor, the probability that a periodogram power
$z$ is above some value $z_0$ is then $P(z > z_0) = e^{-z_0}$
(\citealp{scargle82};~\citeauthor{horne86}~\citeyear{horne86}), 
and, if $M$ independent frequencies are scanned, the probability that
none gives values larger than $z_0$ is $(1-e^{-z_0})^M$. Then, the
{\it false-alarm probability} of the null hypothesis that the data are pure
noise is computed for each value of $z$ as:
\begin{equation}\label{falsealarm}
P( > z_0) = 1 - (1-e^{-z_0})^M
\end{equation}

This expression provides an estimate of the significance of each given
peak in the spectrum that can be identified. For sufficiently high values
of $z_0$ the false-alarm probability is very small, and this corresponds
to the detection of a highly significant periodic signal.

The choice of $M$ is critical in assessing the true significance 
of a given peak in the spectrum. In general, the larger the number of
frequencies, the less significant is any `bump' in the periodogram 
power. The actual value of $M$ depends in general on the 
actual frequency range scanned, and on the number and clumpiness 
of the data points available for the investigation. For this reasons, 
it is difficult to derive a closed analytical form for the
number of independent frequencies. For period
searches between the average Nyquist period ($T_\mathrm{Nyq}\simeq
2L/N_m$) and the duration of the data set a reasonable choice is
$M\simeq N_m$ (\citeauthor{horne86}~\citeyear{horne86}). 
However, when sampling 
very short periods, in general frequencies much higher than the
average Nyquist frequency are reached, then the number of
independent frequencies will be higher. For any particular case,
an actual value of $M$ can be estimated by simple Monte Carlo
analysis: holding fixed the number and epochs of the observations,
generate synthetic datasets containing only Gaussian (normal)
noise representative of the actual errors, then find the largest
value of $P(> z_0)$ and fit the resulting distribution for $M$ in
Eq.~\ref{falsealarm}. By doing so, Marcy \& Benitz~\cite{marcy89} 
and~\citeauthor{cumming99}~\citeyearpar{cumming99} derive $M = 1.15 N_m$ 
and $M = 1.175 N_m$, respectively, and this may be compared with the
analytic relation $M = -6.4 + 1.19 N_m + 0.00098 N_m^2$ by
~\citeauthor{horne86}~\citeyearpar{horne86}. 
Given the illustrative purpose of our
analysis, we rely on results from the literature and choose a typical 
value $M = 1.20 N_m$ throughout our simulations.

\subsection{Procedure for Multiple-Planet Orbital Fits}\label{multifits}

The iterative method for the solution of the non-linear systems of
equations of condition is based on a slightly modified version of
the Levenberg-Marquardt algorithm, as described in S02, which
ensures stability of the solution. It has been upgraded to allow
for the presence of multiple, independent planetary perturbations.

The photocenter motion of the target is described by an analytical model
in which the recomputed relative delay between the target and its $j$-th
reference star ($j=1,\dots,N_r$) is in the form:
\begin{eqnarray}\label{omenc}
  \Delta d_{\star,j} &=&
  \mathbf{B}\cdot(\mathbf{S}_{\star}(\lambda_\star,\,\beta_\star,\,
  \mu_{\lambda,\star},\,\mu_{\beta,\star},\,\pi_\star,\, \sum_{l=1}^{n_p}
(X_{1,l},\,X_{2,l},\,X_{3,l},\,X_{4,l},\,e_l,\,T_l,\,\tau_l)) \nonumber\\
&& -\mathbf{S}_j(\lambda_j,\,\beta_j,\,
  \mu_{\lambda,j},\,\mu_{\beta,j},\,\pi_j))
\end{eqnarray}

The five astrometric parameters (two positions, two proper motion
components, and parallax) of the target ($\lambda_\star,\,
\beta_\star,\,\mu_{\lambda,\star},\,\mu_{\beta,\star},\,\pi_\star$) and
each of the $N_r$ local reference objects ($\lambda_j,\,\beta_j,\,
  \mu_{\lambda,j},\,\mu_{\beta,j},\,\pi_j$) are defined in ecliptic
coordinates (see S02 for details), while for the seven orbital elements
needed to describe the orbit of each of the $n_p$ planets 
around the target we have
utilized the Thiele-Innes representation, in which the semi-major axis
$a_l$, the inclination $i_l$, the longitude of pericenter $\omega_l$ and the
position angle of the line of nodes $\Omega_l$ of the $l$-th planet are
combined to form the quantities:
\begin{eqnarray}
X_\mathrm{1,l} &=& a_l(\cos\omega_l\cos\Omega_l-
\sin\omega_l\sin\Omega_l\cos i_l) \nonumber \\
X_\mathrm{2,l} &=& a_l(-\sin\omega_l\cos\Omega_l-
\cos\omega_l\sin\Omega_l\cos i_l) \nonumber \\
X_\mathrm{3,l} &=& a_l(\cos\omega_l\sin\Omega_l+
\sin\omega_l\cos\Omega_l\cos i_l) \nonumber \\
X_\mathrm{4,l} &=& a_l(-\sin\omega_l\sin\Omega_l+
\cos\omega_l\cos\Omega_l\cos i_l) \nonumber
\end{eqnarray}

In this model, the observation equations are strictly nonlinear
only in the orbital period $T_l$, the eccentricity $e_l$, and the
epoch of pericenter passage $\tau_l$. At the end of each
simulation, the classic orbital parameters are recomputed from the
Thiele-Innes elements~\citep[see for example ][]{heintz78}. The mass
$M_\mathrm{p,l}$ of the $l$-th planet is then determined by
combining the values of $T_l$ (in years), $\pi_\star$ and the 
semi-major axis of the stellar orbit
around the barycenter of the system $a_\mathrm{\star,l}$ (both 
in arcseconds), all estimated from the fit to the
observed delays, and the stellar mass $M_\star$, assumed perfectly
known, in the following approximation of the mass-function formula
for $\mathrm{M}_\mathrm{p,l}\ll \mathrm{M}_\star$ :
\begin{equation}
M_\mathrm{p,l} \simeq \left(\frac{a_{\star,\mathrm{l}}^3}{\pi_\star^3}
\frac{M_\star^2}{T_\mathrm{l}^2}\right)^{1/3} \hskip 1cm
\mathrm{for\,\, l=1,\dots,n_p}
\label{massfunc}
\end{equation}

The multiple-orbit fitting procedure also determines the formal errors on
each estimated parameter $x_m$ as:
\begin{equation}
\sigma(x_m) = \sqrt{C_{mm}},
\end{equation}
where the $C_{mm}$'s are the diagonal elements of the covariance matrix
of the fit~\citep{press92}.
In the case of the orbital parameters recomputed from the
Thiele-Innes elements and the mass of each planet, formal errors are
propagated utilizing the classic error propagation formula, including
correlations between the measured parameters:
\begin{equation}\label{propaga}
\sigma(z_k) = \sqrt{\sum_{i=1}^{m_f}\sum_{j=1}^{m_f}\frac{\partial z_k}
{\partial x_i}\frac{\partial z_k}{\partial x_j}\sigma(x_i)\sigma(x_j)},
\end{equation}
where the sums are extended to the $m_f$ fitted parameters from which
the $k$-th recomputed quantity $z_k$ depends.

As already highlighted in S02, convergence of the non-linear
fitting procedure and quality of the orbital solutions can both be
significantly affected by the choice of the starting guesses. In
the studies presented here this issue has been left aside because,
focusing on the set of presently known planetary systems, we have
utilized as starting guesses the values of the orbital parameters
for the planets in each system published in the literature.
Nevertheless, whenever data will be inspected to search for
planetary signatures in the absence of {\it a priori} information
on the actual presence of planets around a given target, important
questions will have to be addressed and answered, related to how
and to what extent effective starting values, i.e., leading to
successful orbital solutions, will be identified from the data as
a function of actual instrument performances, uncertainties in the
error model, and intrinsic properties of the planetary systems. To
this end, realistic global search strategies will have to be
tested and double-blind test campaigns conducted. Work on these
issues is in progress and will be reported in the future.

\section{Results}

In addressing the problem of the detection and measurement of
multiple planets with SIM, we have utilized the entire set of
existing planetary systems known to date as representative test cases.
The results are discussed below, as follows. First, we study SIM's
efficiency in multiple-planet detection by means of a method that
combines standard $\chi^2$ statistics, periodogram analysis and 
Fourier series expansions. Next, we estimate SIM's ability
to make accurate measurements of multiple-planet orbits and masses.
Then, we gauge the boundaries of SIM's capability to determine the
coplanarity of the orbits of planets in systems. Finally, we quantify
how beneficial it would be to be able to combine astrometric and
spectroscopic orbital elements in order to improve the accuracy in
full-orbit reconstruction, planet mass determination, and coplanarity
measurements. All results reported here have been obtained assuming
end-of-mission analysis of the simulated datasets. We do not discuss
in this paper modifications to the strategy for the analysis of
SIM observations at earlier stages of the mission, when the instrument
operations are still ongoing.

\subsection{Detectability of Multiple-Planet Systems}

In this Section we focus on establishing the
boundaries of reliable multiple-planet system detection with SIM
for a wide range of orbital arrangements, including low values
of the astrometric signal-to-noise ratio $\alpha/\sigma_d$ and
orbital periods exceeding the mission lifetime. To this end, 
we have investigated the performances of various detection
methods ($\chi^2$ test, Fourier series, periodogram analysis) in
the different regimes. Our study is primarily illustrative, and clearly
does not cover all possible arrangements of planetary systems, such as 
configurations with extreme orbital shapes and orientations, and small 
amplitudes of the astrometric signal (either because of small 
planet masses, or due to large distances).

\subsubsection{Probabilities of Detection}\label{probdet}

As discussed in S02, the easiest way to quantify SIM's sensitivity
to planetary perturbations is to parameterize it through the orbital
period $T$ and the {\it
astrometric signal-to-noise ratio} $\alpha/\sigma_d$, where
$\sigma_d$ is the single-measurement error on each relative delay
measurement, while the {\it astrometric signature} $\alpha$ is the
apparent amplitude of the gravitational perturbation induced on
the observed star by a single companion, defined as:
\begin{equation}\label{signa}
  \alpha = \frac{M_\mathrm{p}}{M_\star}\frac{a_\mathrm{p}}{D},
\end{equation}
where $M_\mathrm{p}$, $M_\star$ are the masses of the planet and
star respectively (in solar masses), $a_\mathrm{p}$ the semi-major
axis of the planetary orbit (in AU), $D$ the distance of the
system from the observer (in pc). In the case of multiple,
independent Keplerian perturbations, the orbital motion of the
star about the barycenter will contain components reflecting the
orbital periods of each of the companions, with the peak amplitude
corresponding to the object with the larger value of the 
product $M_\mathrm{p}\times a_\mathrm{p}$ 
(for a given distance of the system and mass of the central star).
It must be noted that Eq.~\ref{signa} constitutes only an upper
limit to the actual magnitude of the measured perturbation, without 
any projection or 
eccentricity effect. If the plane of a circular orbit lies in the
plane of the sky, the two projections of the perturbation will be
equal, with amplitudes given by Eq.~\ref{signa}. If the orbital
plane is inclined with respect to the line of sight, there will
always be some projection of the orbit on the plane of the sky
which has the amplitude given by Eq.~\ref{signa}, but the
amplitude at other positions angles will be less. If the orbit is
eccentric, there could be cases where the orientation of the
orbital plane with respect to the observer's line of sight gives
rise to significantly smaller apparent amplitudes than indicated
in Eq.~\ref{signa}: in the worst-case scenario, when the line of 
apsides is aligned with the line of sight the maximum
measured amplitude corresponds to the orbital semi-minor axis,
which is still equal to $> 0.95\, a_\mathrm{p}$ for moderately
eccentric orbits ($e < 0.3$), but can drop below $0.5\,
a_\mathrm{p}$ for high eccentricities ($e > 0.8$). For very
eccentric orbits, the projected amplitude of the stellar orbital
motion can thus be up to a factor 2 or more smaller than the value
calculated with Eq.~\ref{signa}. The variety of orbital
arrangements reproduced by the set of planetary systems known to
date does not present any such case, as the eccentricity of the
orbits is usually moderate ($e \leq 0.55$). The only exception is 
HD 160691 b ($e = 0.8$), but its astrometric signature 
($\alpha = 139/\sin i$ $\mu$as) is sufficiently high not 
to constitute a problem in terms
of ability to detect it or to determine its orbit, even in the
worst-case scenario.

In order to tackle the issues discussed in the previous Sections,
we ran simulations of SIM narrow angle astrometric campaigns
dedicated to the observations of the 11 stars known to harbor
multiple-planet systems, utilizing the observing scenario sketched
in Section~\ref{scenario}. In particular, we generated sets of
500 planetary systems on the celestial sphere. The values we 
utilized for the stellar parameters and orbital elements were taken 
from various online catalogs in September 2002, as summarized in 
Table~\ref{planparam}.\footnote{For an up-to-date list of 
extra-solar multiple-planet systems and their orbital elements as 
they are known at the time of writing see for example Fischer et al. 
(2003)} The analysis carried out and
results presented in this and the following Sections are based on
a total of 990\,000 simulated planetary systems.

Two orbital elements cannot be determined by radial velocity
measurements, namely $i$ and $\Omega$. We simulate systems with
random uniform distributions for the lines of nodes
($0^\circ\leq\Omega\leq 180^\circ$) and express detection
probabilities as a function of the orbital inclination, 
constrained to vary in the range $1^\circ\leq i \leq
90^\circ$. Planet masses are
scaled with the co-secant of the inclination angle.

An important point is the choice of the relative inclination angle
between pairs
of components in multiple systems. The relative inclination
$i_\mathrm{rel}$ of two orbits is defined as the angle between the 
two orbital planes, and is given by the formula (see for 
example Kells et al. 1942):
\begin{equation}\label{inclrel}
\cos i_\mathrm{rel} = \cos i_\mathrm{in}\cos i_\mathrm{out}+ \sin
i_\mathrm{in}\sin i_\mathrm{out} \cos(\Omega_\mathrm{out}-
\Omega_\mathrm{in}),
\end{equation}
where $i_\mathrm{in}$ and $i_\mathrm{out}$, $\Omega_\mathrm{in}$
and $\Omega_\mathrm{out}$ are the inclinations and lines of nodes
of the inner and outer planet, respectively. We choose
the orbits of the multiple-planet systems shown in
Table~\ref{planparam} to be perfectly coplanar, i.e. $i$ and
$\Omega$ are exactly identical for all the components in each
systems. This choice has been adopted everywhere, unless otherwise
noted.

There are three important questions related to the use of $\chi^2$
statistics as a tool for multiple-planet detection that we aimed
to answer: 
\begin{itemize}
\item[a)] does detection probability depend on the shape and
orientation of the orbits?\footnote{In previous works 
(\citealp{lattanzi00}; S02) 
it was found that, for the case of single planets,
averaging over orbital eccentricity, the astrometric detection
probability based on the $\chi^2$ test of the null hypothesis that
the star is single did not depend upon the apparent orientation of
the orbit} 
\item[b)] is detection of low signal-to-noise astrometric
perturbations (the detection limit intrinsic to the $\chi^2$ test
is reached for $\alpha/\sigma_d\rightarrow 1$) influenced by the
previous removal of larger astrometric signals? 
\item[3)] what is the
behavior of false detections (once all planetary perturbations
have been detected and removed) as a function of a broad range of
orbital arrangements and astrometric signals?
\end{itemize}
Figure~\ref{fig1} shows, for each planet in each of the 11 systems
considered, the results of the $\chi^2$ test expressed as a function 
of the orbital inclination $i$. For each given system, the test was 
applied to the residuals to a single-star model, to a model containing the
previously detected companion(s), and finally after the removal of all the
Keplerian motions. In each system, the dominant signature is always
easily detectable, regardless of the inclination of the orbital
plane. Even long-period planets such as 55 Cnc d, due to its large
mass, are unambiguously discovered. When the companion producing
the larger signal is removed, the remaining components, due to the
usually large astrometric perturbation induced on its parent star
(even for edge-on configurations), are also detected without
ambiguities, independent of the value of $i$, with the 
exception of the inner planet around HD 38529,
the innermost planet orbiting $\upsilon$ And, and  the
middle planet in the 55 Cnc system.\footnote{We recall
that the astrometric signature $\alpha\propto T^{2/3}$, while
the radial velocity amplitude $K\propto T^{-1/3}$, thus the
{\it Hot Jupiters} on very short-period orbits, the first to
be detected by spectroscopy, are most difficult to detect with
the astrometric method} For the first two objects,
in fact, in the edge-on configuration, the amplitude of the
astrometric perturbation is of order of $\sigma_d$, thus the $\chi^2$
test fails to recognize their presence about sixty percent of the
time. 55 Cnc c has a minimum signature $\alpha\simeq 4$ $\mu$as,
yet its detection probability in the edge-on configuration is
a little below the 95\% detection threshold due to the fact that
its minimum mass also corresponds to the configuration in which
only one dimension of the orbit is measurable by astrometry.
As $i\rightarrow 0^\circ$ , towards a face-on configuration, their
projected masses increase, and in all cases detection probability
reaches 100\% for a value of $\sin i$ such that the correspondent
astrometric signature $\alpha\simeq 2\sigma_d$. This result
confirms what had been found in our previous works 
(Casertano et al. 1996; Casertano \& Sozzetti 1999; 
\citealp{lattanzi00};~\citeauthor{sozzetti01}~\citeyear{sozzetti01}; 
S02). In particular, we find that,
as long as the orbital period does not exceed the mission
lifetime, detectability of astrometric perturbations close to the
detection limit is not hampered by the presence of one or more
companions producing larger signals. After the larger
perturbations have been fitted and removed, a signal-to-noise
ratio $\alpha/\sigma_d\simeq 2$ is sufficient for secure detection
(at the 95\% confidence level), just as if the planet was the only
orbiting companion.

The last point addressed in Figure~\ref{fig1} concerns the fraction of false
detections after all the planets have been discovered and removed.
Due to the confidence level adopted for the $\chi^2$ test, we
expect $\sim 5\%$ of false positives if the subtraction of all
planetary signatures has been successful. On the other hand,
we may expect in principle that orbits with long or very short
periods and producing low signal-to-noise ratios might not be
accurately fitted and removed, with the result of larger spurious
residuals, that could translate in a larger amount of false
detections. As shown in Figure~\ref{fig1}, the rate of false positives is
roughly independent on the value of $i$, and comparable for all the systems
investigated. Again, three exceptions are present, 55 Cnc c, HD 38529 b,
and $\upsilon$ And b. In these three cases, the signal to be fitted 
approaches the detection limit as we move
towards $i = 90^\circ$. The quality of the fits when $\alpha\simeq
\sigma_d$ is poor, and this degradation translates into a
number of false detections up to a factor 2 larger than the expected
5\%, with a more pronounced trend as we move close to edge-on orbits.
Instead, whenever all the planets in a given system produce a signal
$\alpha > 2\,\sigma_d$ (even in the minimum mass configuration),
then the number of false detections drops even below the expected
threshold. We find that the average false
detection rate, including all systems, is $\overline{F_d} = 3.2\%$, 
well within the predicted 5\% of false positives in case
of satisfactory removal of all planetary perturbations.

\subsubsection{Periodogram Analysis}\label{periods}

The $\chi^2$ test only allows us to draw general conclusions on the
need to reject or keep an a priori theoretical model describing the
observations, but provides no clues to the actual nature of the
residuals, if the assumed model is proven to be not consistent with
the data. In order to provide a direct identification of the
nature of the observation residuals and determine the true significance
of planet detection/non-detection, a standard procedure is that of
inspecting the data by means of a periodogram analysis, to search for
evidence of the presence of a clear periodicity in the time-series.

At the end of each $\chi^2$ test, we perform a periodogram analysis
(utilizing the tools described in Section~\ref{stattools}) to assess
the presence or absence of further periodic signals. This test is
applied to a set of observations that contain an a priori known set
of planetary signatures, but nonetheless it has proven a useful
exercise in order to gauge the ranges of applicability of this procedure
to future, actual observations.

Table~\ref{scargle} shows a set of results pertaining to simulations
of the 11 planetary systems assuming an average value of the common
inclination angle ($i = 45^\circ$). The crucial points to be discussed
are related to how well the period of the signal can be identified,
and to how strong the signal must be with respect to the single-measurement
precision, in order to give a peak $z_p$ in the periodogram power
allowing for a significant detection, i.e. a small value of the false-alarm
probability. For the purpose of this analysis, we set a threshold
$P(z_p) \leq 10^{-3}$, which implies, given the assumed number of
independent frequencies discussed in Section~\ref{stattools}, a value
of $z_p\geq 12.0$ to ensure a significant detection of a periodic
signal.

From inspection of Table~\ref{scargle}, the first conclusion that can
be drawn is that even in the case of an astrometric signal-to-noise
ratio $\alpha/\sigma_d\simeq 1$, as long as the orbital period is
shorter than the mission duration, a significant peak can be found in
the spectrum. This is the case for example for HD 38529 b, that exhibits
$\alpha/\sigma_d = 1.23$ (for $i= 45^\circ$), and while $\chi^2_\nu$
approaches 1 (in fact, from Figure~\ref{fig1} its presence goes undetected
about 15\% of the time), the false alarm probability is well below the
threshold, and the actual peak in the spectrum matches the true orbital
period to within $\sim 1\%$. A graphical interpretation of the results 
for this system is given in Figure~\ref{fig2}.

Secondly, three main period ranges can
be identified, over which the efficacy of the periodogram analysis
is significantly different. First of all, the orbital period may happen
to be short enough that the spectrum may peak at a value of $T$
which departs significantly from the true one. This is for example the case
of $\upsilon$ And b, for which (see Figure~\ref{fig1}) the probability of
detection is close to 100\% (in fact, $\alpha/\sigma_d = 1.64$, at
$i = 45^\circ$), but the period is more than a factor 3 shorter than that
of HD 38529 b, and the periodogram analysis fails to recognize the correct 
value of $T$, identifying a significant peak in the spectrum at a period
value which is about an order of magnitude off the true one. As shown
in Figure~\ref{fig3}, bottom-left panel, 
only the second highest peak is close to the
actual value of the period of $\upsilon$ And b, but still about a factor
1.5 off. The failure
of the periodogram analysis at very short periods can be dealt with at
least in part in two ways. First of all, this problem may be 
corrected by increasing the number of observations.
From the results in Table~\ref{scargle}, in fact, we see how, in the
case of randomly spaced data, it is possible to retrieve meaningful
information on periods much shorter than the `average' sampling rate
in the case of evenly spaced data. For 24 two-dimensional observations
distributed over 5 years, the average sampling rate would be $\sim 0.2$
yr, or 72 days. Due to the fact that {\it some} points are spaced much
closer than that, the periodogram can effectively identify signals
with periods as short as 15 days. Increasing the number of observations
would obviously increase the ability to recover even shorter periods.
Improvements in this direction can be obtained also by extending
the range of test frequencies to values that are much greater than
the average Nyquist frequency, and by over-sampling the spanned
frequency range, in order to obtain information on frequency bins that are
much smaller than the typically lowest independent frequency, that
corresponds to the inverse of the timespan of the input data (i.e., the
frequency such that the data can include one complete cycle). In 
our present analysis we have not quantified the effect of these 
different choices. 

Furthermore, the results summarized in Table~\ref{scargle} tell us that
well-sampled periods up to roughly the mission lifetime
constitute `easy' cases,
for which the correct period can be identified with high confidence and
accuracy (typically better than 7-8\%), and this independently on the
actual magnitude of the astrometric signature, as far as it lies above
the single-measurement error. Figure~\ref{fig4} provides a graphical
example of the results in Table~\ref{scargle} for the `easy' case
of HD 82943.

Finally, when $T > 5$ yr we enter a regime in which the classic
periodogram is intrinsically doomed to fail. Increasing the number
of observations (without increasing the mission length) or
manipulating the frequency range and the way it is sampled does
not improve the results. In fact, several authors (e.g.,
~\citealp{black82};~\citealp{walker95};~\citeauthor{cumming99}
\citeyear{cumming99}) in the past highlighted that the periodogram 
power at periods exceeding the timespan of the data can be 
significantly reduced in the traditional periodogram formula.
From the results in Table~\ref{scargle} we note how, in all the
cases where the actual orbital period of the planet is greater
than 5 years, the periodogram still identifies a significant
periodic signal, but the peak systematically appears at a period
of order of the total timespan of the data. Figure~\ref{fig5}
provides a graphical representation of the data in Table~\ref{scargle}
for the classic case of 55 Cnc, which harbors the planet with the
longest period discovered so far. The situation is somewhat
different from the spectroscopic case,where as $T$ lengthens 
the radial velocity amplitude decreases, and indeed the peaks in
the power spectrum relative to long period signals are
significantly reduced. For astrometry, the amplitude of the signal
increases with increasing $T$, and in the sample of the presently
known extra-solar planetary systems all companions are massive (as
they are easier to detect by spectroscopy). The astrometric
signal-to-noise ratio $\alpha/\sigma_d\gg 1$, and so, even if only
a fraction of the orbit is sampled during the 5 years of simulated
mission lifetime, the acceleration in the residuals is easily
disentangled from the proper motion and a significant peak in the
spectrum can also be identified. Nevertheless, the exact period
cannot be recovered in a periodogram analysis. 
This systematic trend of a significant peak
detected at $T\simeq L$ whenever the true period is greater than
the timespan of the data is likely to become less evident for 
sufficiently small-mass planets (or, equivalently, sufficiently
distant systems) on sufficiently long-period orbits. In fact, a
reduction in the measured amplitude of long-period signals may
occur either in the case of $\alpha/\sigma_d\rightarrow 1$ or when
$T$ is so long that secular effects such as proper motion begin
overlapping efficiently to the astrometric signature. We have not
investigated these issues here, as the main focus of this paper is
on the actual characteristics of the multiple-planet systems
discovered so far, and how they can be detected and measured with
SIM.

\subsubsection{Search for Periodicities when $T > L$}\label{fourier}

As we have seen, there exist regimes of orbital periods (shorter than
the SIM mission lifetime) for which the Lomb-Scargle periodogram
approach is robust, but others (comparable to
or longer than the duration of the observations) for which the
periodogram analysis fails to provide sensible results.
One possible solution to overcome the `impasse' was proposed
in the recent past (\citealp{walker95};~\citealp{nelson98};
~\citeauthor{cumming99}~\citeyear{cumming99}). It 
consists of utilizing the zero point of the sinusoid in the fit as a
free parameter, i.e. letting the mean of the data to float during the fit.
This is contrary to what happens in the classic periodogram formula, in
which the mean (taken as a proxy for the zero point)
is subtracted from the data, and thus assumed to be perfectly known.
When applied to radial velocity data, the floating mean periodogram
approach has been shown to be successful at
confirming existing companions and characterizing variations in the
observation residuals due to long-period companions with orbits still
awaiting phase closure~\citep[e.g., ][]{fischer01}. We choose a different
approach, with the aim of illustrating the potential of a strategy for
period searches that tiles the classic periodogram approach to
the Fourier analysis of periodic signals. When combined with the standard
$\chi^2$ statistics, this multi-method approach to planet detection
provides an operational framework that may also be applied more in
general to datasets from other astrometric observatories such as
the upcoming ESA Cornerstone Mission GAIA~\citep{perryman01}.

From the theory of Fourier analysis we know that a periodic, but
non-sinusoidal, motion can be represented by a series containing a
sinusoidal term at the fundamental frequency (corresponding to the actual
period of the motion) plus higher order harmonics at exact multiple
integers of this frequency, which carry on the information about the
magnitude of the distortion from a purely sinusoidal motion due to the
eccentricity (see for example Monet 1983).
In the tangent plane to the celestial sphere, for the $j$-th planet in a
system, the two components of its
projected orbital motion can then be written as:
\begin{eqnarray}
x_{j,tan} &=& \sum^{\infty}_{k=0}a_{k,j}\cos k\gamma_j t +
\sum^{\infty}_{k=0}b_{k,j}\sin k\gamma_j t \nonumber \\
y_{j,tan} &=& \sum^{\infty}_{k=0}c_{k,j}\cos k\gamma_j t +
\sum^{\infty}_{k=0}d_{k,j}\sin k\gamma_j t \nonumber
\end{eqnarray}
where $j=1,\dots,n_p$, and $\gamma_j = 2\pi/T_j$.
The series is quickly convergent except for
large eccentricities. For moderately eccentric orbits, say $e\leq 0.5$,
three terms (fundamental, first and second harmonic) are sufficient to
cover more than 95\% of the signal~\citep{jensen73,konacki02}. 
For the purpose of our analysis, we have
tested the ability of the above model to detect a significant long-period
signal in the case of the five multiple systems
(HD 37124, HD 38529, HD 74156, 55 Cnc, and 47 UMa)
containing one planet on a moderately eccentric ($e\leq 0.40$),
long-period orbit ($T > 5$ yr). We have limited the Fourier expansion
to the second harmonic term ($k_{max} = 3$), and performed a linear
fit (the unknowns being the parameters $a_k$, $b_k$, $c_k$, $d_k$, for
$k=0,\dots,k_{max}$) to the simulated SIM astrometric data relative to the
systems described above, as a function
of a large number of trial frequencies (4000) corresponding to the
period range 0.01-20 yr. Under the assumption of coplanarity of the orbits,
then in the case of 47 UMa the inner planet has a
larger signature, thus the outer planet signal has been searched for
utilizing a model in which the Fourier expansion is superposed to
a fully Keplerian orbit for the inner planet. The results are summarized
in Figure~\ref{fig6}.

The two upper, the two middle, and the lower left panels show
the behavior of the reduced chi-square $\chi^2_\nu$ as a function of the
dense grid of trial periods for the abovementioned systems, again 
setting an average value of the inclination $i = 45^\circ$. As opposed
to the periodogram analysis, the Fourier series approach is capable
of identifying the correct period of the outer planet, corresponding to 
the minimum of $\chi^2_\nu$, to better than $4\%$ in all the five 
cases. The 
periodicity with the larger amplitude can be correctly retrieved with
high accuracy independently of the underlying presence of the second,
inner companion, which is clearly evident from the fact that $\chi^2_\nu$
still departs significantly from unity even at its minimum. There are
two exceptions, i.e. 47 UMa c, beyond which at present no other planetary
companions have been discovered,
and HD 38529 b, for which $\chi^2_\nu$ appears to be consistent with the
outer planet being the only one in the system due to the very low
signature induced by the inner planet. Nonetheless, as we have
seen in the previous Section, the signal from the inner companion can
still be identified in the periodogram power spectrum with high
confidence, even in a case in which the $\chi^2$ statistics approach
is more sensitive to failure. These results underline the importance of
utilizing a range of different tools for planet detection, which can be
combined in order to improve significantly the efficiency with which
planetary signatures can be revealed even at low values of the astrometric
signal-to-noise ratio $\alpha/\sigma_d$.

Finally, we note how in Figure~\ref{fig6} the minima of $\chi^2_\nu$ 
for HD 37124 c, 55 Cnc d, and 47 UMa c are broader than those for 
HD 74156 c and HD 38529 d. This is primarily due to the combined 
effect of the magnitude of the astrometric signature induced on the 
parent star and the fraction of the orbit covered by the observations. 
For comparison, the
lower right panel shows the results of a simulation in which
the same fitting procedure was applied to our own
solar system (nine independent, coplanar Keplerian orbits),
placed at different distances from the observer. The
dominant signature is the one associated with Jupiter, and its period
is easily identified with an accuracy of 5-6\%. As we move farther
away from the observer, the value of the minimum of $\chi^2_\nu$,
which is initially located at $\chi^2_\nu(min) =  1.26$, indicating
the presence of a second underlying signal (from Saturn, the second
heaviest planet in our system and the one producing the second largest
signature), decreases, and eventually at 10 pc it is
perfectly consistent with Jupiter being the only planet around the Sun.
At the same time, while the presence of the second largest planet becomes
undetectable, also the signature from Jupiter decreases and
correspondingly, for the same fraction of the orbit covered by the
observations, the slope of the $\chi^2_\nu$ curve at long periods
becomes increasingly less steep. 

In the perspective of the
optimization of a global search strategy for the determination of the
best configuration of starting values for the orbital parameters in the
fully Keplerian fit {\it for newly discovered planetary mass objects}, 
a precise guess for the actual value of the orbital 
period is needed, in order to try to minimize the dimensions of the 
parametric space to be searched. Suppose in fact we select as plausible 
preliminary values of $T$ those whose $\chi^2_\nu$ difference with 
the minimum $\chi^2_\mathrm{\nu,min}$ is less than 15\% of the latter 
value, as done by~\citeauthor{guirado97}~\citeyearpar{guirado97} 
in their astrometric orbit 
determination of a low-mass companion around the star AB Doradus. 
This would mean that, for the cases shown in Figure~\ref{fig6}, 
assuming no spectroscopic orbits are available, 
plausible starting guesses for the orbital period in a subsequent 
global least-squares fit would lie in the ranges reported in 
Table~\ref{plausible}. As it is clear, 
whenever we have a configuration approaching 
a regime in which $\chi^2_\nu$ has a very broad minimum (HD 37124 c, 
47 UMa c, 55 Cnc d, Jupiter at 10 pc), this corresponds 
to a range of plausible starting guesses for the period covering 
several years, and it is likely this would cause accurate orbit 
determination and mass measurements to become a more difficult task. 

Clearly, the results shown in these Sections constitute only a very 
preliminary investigation of the potential of the approach to 
the problem of real-life detection of {\it newly discovered planetary 
systems} with a method that combines $\chi^2$ statistics, periodogram
and Fourier analysis of SIM astrometric observation residuals. Due to
the illustrative purpose of our work, some important topics have been left
aside, such as estimates of the degradation in SIM's ability to detect
planets with low values of $\alpha/\sigma_d$ on long-period orbits due
to significant correlation with the target's proper motion. Also, some 
relevant technical issues remain open, such as the possibility to 
not only provide robust estimates of the reliability of the adopted 
model, through additional statistical 
hypothesis testing~\citep[e.g., ][]{bevington03}, but also assess the 
reliability of the estimates themselves, through 
further testing and placing statistical constraints on the post-fit 
observation residuals and their covariance matrices, 
in order to select the optimal least-squares solution
~\citep[e.g., ][]{bard74,bernstein97}. Nevertheless, 
within the limited scope of our study, we have been able to provide
quantitative examples which are 
useful for understanding that it will be very beneficial if a variety of
numerical and statistical tools will be implemented to maximize the
robustness of the method aimed at verifying the true significance and
reliability of a detection.

\subsection{Multiple-Planet Orbits Reconstruction}

Once a reliable detection of a system of planets has been established,
a number of other important issues must be addressed, all primarily
related to how well it is possible to measure the orbital parameters
and masses of each planet.

As soon as the number of planets in a system is greater than one, the
parameter space to investigate becomes immediately very large. In this
illustrative study, we have
parameterized the ability of SIM to measure multiple-planet systems
in terms of their astrometric signatures, orbital periods,
eccentricities, and inclinations of the orbital planes. We have then
investigated the potential of SIM to make meaningful coplanarity
measurements by gauging how accurately the relative inclination of pairs
of planetary orbits may be determined. Finally, we have quantified the
improvement in the accuracy on the determination of orbital parameters
and masses that maybe reached when both astrometric and spectroscopic 
data are available.

\subsubsection{Accuracy in Orbit and Mass Determination}\label{reconstruct}

The first relevant question to be addressed is: what is the impact
on the accuracy in orbital parameters and mass determination when the
number of planets in a system is larger than one, and the number of
parameters to be fitted correspondingly increases?

Table~\ref{ratios} shows, for each planet in the 11 systems under
consideration, the ratios of rms errors $\sigma_m$
for the most relevant parameters ($M_p$, $a$, $T$, $e$, and $i$)
derived from a multiple-planet fit to the same quantities
$\sigma_s$ obtained from a single-planet solution (in simulations
where only one planet was generated around the observed parent star).
The values are averaged over the inclination angle. 

In general, for planets with periods $T \leq L$,
the two-planet fit (or three-planet fit, as in the case of
$\upsilon$ And and 55 Cnc) degrades the orbital elements and mass 
estimates by $\sim 30-40$ \% with respect to the single-planet
fit, with typical fluctuations of order of $5-10$ \% among different
parameters. From closer inspection of Table~\ref{ratios},
some particularly interesting cases stand out.

First of all, there is a clear trend for larger values of the ratio
$\sigma_m/\sigma_s$ for a given parameter, as the magnitude of the 
astrometric signature decreases. For example, HD 82943 b and HD 82943 c,
with signatures of only $65/\sin i$ $\mu$as and $22/\sin i$ $\mu$as
respectively (see Table~\ref{planparam}),
have values of the rms errors typically about a factor 2 or more larger
than the respective single-planet case, as opposed for example to the
case of the Gliese 876 system, in which the two planets, producing larger
signals, are measured together only about 20\% worse than if they were
the only orbiting bodies.

The second relevant feature arising from the results of
Table~\ref{ratios} concerns those systems in which the orbit of
the outer planet is not fully sampled during the 5 years of
simulated SIM observation. In three such cases (HD 38529 c, HD
37124 c, and HD 168443 c), the ratios $\sigma_m/\sigma_s\leq 1$
for almost all parameters shown in Table~\ref{ratios}. The reason
for this is to be found in the intrinsic superiority of the model 
containing more than one planet, when it is applied to configurations 
where the outer planet has $T > L$ and produces a signature much 
larger than the inner one. In fact, the larger number of free 
parameters (with respect to the single-planet case) is compensated 
by the superior model, and 
the net result is that in the least squares solution a smaller
fraction of the outer planet's signal is identified as proper
motion, with the consequence of a more precise 
determination of the outer planets'
orbital elements and masses as opposed to the single-planet
solution (up to about a factor 2-3 better). This behavior is not
very prominent for 55 Cnc d, due to its much longer period 
($T\simeq 3L$), which translates in such 
poor sampling of the orbit that the larger signature is not
enough for the orbital solution to improve significantly.

Finally, the case of the 47 UMa system stands out. Assuming
coplanarity of the two planets' orbits, then the astrometric
signature of the outer one is smaller. The inner planet's orbit
is sampled completely only once during the timespan of SIM
observations, and the outer planet's period is 1.42 times longer.
Even utilizing the spectroscopic orbital elements as starting
guesses for the iterative least square procedure, the net result
is that a significant portion of the inner planet's signal is
assigned to the outer one, and the accuracy in the determination 
of semi-major axis, period, and mass 
is degraded by over an order of magnitude
with respect to the single-planet solution.
For what concerns the uncertainties on eccentricity 
and inclination angle, the degradation effects due to this particular 
configuration are only marginal. However, the orbital solution for 
the outer planet does not benefit from 
this redistribution of the inner planet's signal, and its orbital
parameters and mass are also more poorly determined than in the
single-planet case.

In Figure~\ref{fig7} and~\ref{fig8} we provide a detailed
visualization of the results on orbit reconstruction and planet
mass determination for the three-planet system 55 Cnc and the
two-planet system HD 12661, which are representative of the
variety of configurations addressed in this study, i.e.
well-sampled, well-spaced orbits, or poorly sampled due to either
very short or very long periods, high values of the astrometric
signal-to-noise ratio as well as values of $\alpha/\sigma_d$
approaching 1. In the two Figures, the fractional deviation of the
measured from the true value for the five relevant parameters
discussed above is plotted against the inclination angle, both in
the case of single- and multiple-planet realizations. Four of the
five parameters ($M_p$, $a$, $e$, and $i$) are primarily sensitive
to the value of $\alpha/\sigma_d$. The larger the astrometric
signal-to-noise ratio, the more accurate their measured value,
independently on the orbital period. So for example in the 55 Cnc
system the outermost planet has these parameters best measured
(even though only one third of its orbit is sampled), while the
middle planet's parameters are the least accurately determined. An 
important exception in this case is the orbital eccentricity of the
innermost planet: its orbit is tidally circularized ($e = 0.02$),
and although its astrometric signature is a factor 2 larger than
the middle planet, its period is so short that the inadequate 
sampling causes the value of $e$ to be more poorly constrained as
compared to the one of the middle planet. On the contrary, the
accuracy with which $T$ itself is retrieved primarily depends on
how well the orbit is sampled (center left panels in
Figures~\ref{fig7} and~\ref{fig8}). Thus, wider orbits have the
period less accurately measured, while the accuracy on the
determination of the other parameters is improved.

As the inclination
$i$ decreases, we move from a perfectly edge-on to almost
face-on configuration, i.e. from a situation in which only one
dimension is actually measured to one in which the SIM measurements
provide full two-dimensional information. At the same time, the true mass
inferred for each planet from the radial velocity data increases,
therefore the astrometric signature is larger. These effects combined
translate into increasingly more
accurate measurements of the planet's mass, semi-major axis,
eccentricity, and period, as $i\rightarrow 0^\circ$. At the same time, the
observations are less sensitive to the inclination itself, and its
fractional measurement error increases (lower left panels in both
Figure~\ref{fig7} and~\ref{fig8}).

The present set of extra-solar planetary systems covers a relatively
wide range of shapes and sizes of the orbits, and planetary masses.
We can then attempt to generalize the results presented here to
characterize the variety of configurations of potential planetary systems
in the solar neighborhood for which SIM could provide accurate
measurements, for the given template observing strategy outlined in
Section~\ref{scenario}.
In the three-dimensional parameter space defined by astrometric
signal-to-noise ratio $\alpha/\sigma_d$, period $T$, and eccentricity $e$
we have identified the loci for the measurement of a given orbital
parameter or mass of a planet in a multiple system to a specified
level of accuracy. Figure~\ref{fig9} shows iso-accuracy contours 
for M$_\mathrm{p}$, $a$, $T$, $e$, $i$, and $\Omega$ 
(expressed in percentages of the true value) in the $e-T$ 
plane. The results are averaged over the inclination angle.
The general indication is that the correlation between these
two parameters is relatively limited. When the sampling is poor due to
very short-period orbits, the degradation in accuracy is common to all
parameters, and regardless of the orbital eccentricity. For well-sampled
orbits, the parameters can be measured better, with
varying degrees of improved accuracy. Again, the results are essentially
insensitive to $e$. In fact, massive planets on relatively long-period 
orbits such as 55 Cnc d produce astrometric signals large enough that
the fraction of the orbit covered by the observations still allows for
accurate measurements of the orbital elements, regardless of the
value of the eccentricity. An important exception is constituted
by the accuracy on the eccentricity itself (central right panel),
which appears to degrade for periods exceeding the 5-year
SIM mission duration, unless $e\geq 0.1$.
\footnote{When the orbital period is sufficiently
long or the mass of the planet sufficiently low,
instead, $e$ is likely to play an important role in the
degradation of the reachable accuracy on other parameters as well,
as eccentric
orbits not sampled in correspondence of the pericenter passage may not
be properly measured by the fitting algorithms due to a lack of
curvature sufficient to disentangle the periodic motion from the stellar
proper motion} Finally, it must be noted how the accuracy in the
determination of the orbital period
(central left panel) shows a behavior opposite to that of the
other parameters, in agreement with the fact that how well 
the periodicity of the astrometric signal can be measured 
depends upon the number of times an orbit is
fully sampled. The same mostly uncorrelated behavior is evident in
the plane defined by signal-to-noise ratio and eccentricity.

Figure~\ref{fig10} shows the same results, but this time in the
plane defined by $\alpha/\sigma_d$ and $T$. Here we appreciate a
clear correlation between these two crucial parameters, as it is
shown by the trend for lower accuracy on M$_p$, $a$, $e$, $i$, and
$\Omega$ towards low values of $\alpha/\sigma_d$ and short
periods, and for improved determination of orbital elements and
masses for large values of $\alpha/\sigma_d$ and longer periods.
Again, the orbital period itself behaves in an opposite fashion.

It is possible to provide an analytic representation for the 
dependence of the fractional errors on the 
estimated orbital elements and masses as a function of the three 
quantities $\alpha/\sigma_d$, $T$, and $e$. In order to 
do so, we have performed 
a fit to the data (i.e., the fractional uncertainties 
$\sigma_\mathrm{p}$ on a given parameter, where the subscript 
$\mathrm{p}$ can be equal to $M_\mathrm{p}$, $a$, $T$, $e$, $i$, 
or $\Omega$) with a model of the form:
\begin{equation}
\sigma_\mathrm{p} (\%) = \eta_0\times(\alpha/\sigma_d)^{\eta_1}
			 \times T^{\eta_2}\times e^{\eta_3}
\end{equation}
The results are summarized in Table~\ref{powerlaw}, and are averaged 
over the inclination angle. Again, similarly 
to what we saw in Figures~\ref{fig9} and~\ref{fig10}, we realize 
how, for example, the fractional errors on $M_\mathrm{p}$, $a$, and 
$i$ are dominated by the behavior of $\alpha/\sigma_d$ and 
$T$, while the dependence on $e$ in such cases is relatively weak. 
For $\sigma_\mathrm{e}$, instead, the dependence on the eccentricity 
itself is much stronger. Finally, in the case of $\sigma_\mathrm{T}$, 
the exponent of $\alpha/\sigma_d$, that would have the 
fractional error on the orbital period decrease as the astrometric 
signal-to-noise ratio increases, is largely compensated by the opposite, 
and stronger, 
dependence on $T$ itself, with the result that, as $T$ increases, 
$\sigma_\mathrm{T}$ grows larger as well. It is important to stress 
the limits of applicability of such empirical multi-dimensional 
power-laws. In fact, these formulas do not include a variety of effects, 
such as the degradation induced by the increasingly poor 
sampling at very short or very long periods. Due to the former effect, 
for example, the best-fit formula for $\sigma_\mathrm{T}$ systematically 
underestimates the errors on the orbital period for $T < 0.1$ yr. 
In general, the agreement between the 
power-laws shown in Table~\ref{powerlaw} and the data is good when the 
values of $\sigma_\mathrm{p}$ are larger than $1\%$, while discrepancy 
factors of a few typically 
arise when the fractional errors on the parameters become very small. 
Finally, we expect the functional dependence of the 
fractional errors on $\alpha/\sigma_d$, $T$, and $e$, as well as the 
values of the zero-points $\eta_0$, to be somewhat 
different when a range of possible observing strategies and different 
values of the measurement errors were to be considered. 
In our exploratory work we have not investigated further such issues.

Taking into account the results reported in
Tables~\ref{ratios} and~\ref{powerlaw} and those summarized in 
Figures~\ref{fig9} and~\ref{fig10}, we can attempt to draw some general
conclusions as for what concerns the ability of SIM to measure the
most relevant parameters in multiple-planet systems: on average,
10\% accuracy, or better, will be reached in the determination of
the mass of planets in two or three-planet systems with components
on periods not shorter than $T\simeq 0.1$ yr, and producing an
astrometric signal-to-noise ratio $\alpha/\sigma_d\geq 5$. Similar
accuracy will be attainable for systems containing 
massive planets with periods 
longer than the SIM mission duration (up to $T\simeq 3 L$), which
produce $\alpha/\sigma_d\geq 100$. The semi-major axis, orbital
inclination, and lines of nodes will behave similarly, while the
eccentricity will require
$\alpha/\sigma_d\simeq 10$ for periods $T\leq L$, and accurate
measurements will become increasingly more difficult to achieve 
for periods exceeding the mission lifetime. The orbital period
will be recovered with typical accuracies of a fraction of a
percent for $T\leq L$. Even for objects with $T\simeq 3 L$ and 
producing $\alpha/\sigma_d\gg 1$, such as 55 Cnc d, 1\% accuracy 
should be attainable.

\subsubsection{Coplanarity Measurements}\label{coplan}

As discussed in the Introduction, the increasing number of
extra-solar planetary systems has motivated detailed
theoretical studies on their dynamical evolution and long-term
stability. It is worth mentioning in detail the main results
derived so far for some of the systems which have been the target
of the larger theoretical efforts.
For example, in the case of the $\upsilon$ And system,
neglecting to first order the effects of
the innermost planet on the overall stability of the system,
Stepinski et al. (2000) come to the conclusion that
dynamical stability requires the orbital inclination of the outer
two companions to be greater than $i\sim 13^\circ$, otherwise the
two objects would be too massive and gravitational interactions
would disrupt the system. Furthermore, the system cannot be stable
in the long term if relative inclinations are greater than
$55^\circ$, $35^\circ$, and $10^\circ$, for $i\sim 64^\circ$,
$i\sim 30^\circ$, and $i\sim 15^\circ$, respectively: then, the
more massive the planets, the closer to coplanarity their orbits
have to be, for the system not to be destabilized on a short
time-scale. As for what concerns the strongly interacting
two-planet system on short-period orbits
around Gliese 876, Laughlin \& Chambers (2001),
Rivera \& Lissauer (2001b), and  
Go\'zdziewski et al. (2002) have
found limits for dynamical stability on their relative inclinations
that, similarly to the $\upsilon$ And case, are a strong function
of the unknown inclination of the orbital planes (and thus unknown
masses): the relative 
inclinations of stable systems can vary in the ranges of $\pm 15^\circ$
for $\sin i = 0.5$ up to $\pm 90^\circ$ for $\sin i = 1$.
Finally, Laughlin et al. (2002) have studied
extensively the 47 UMa system, and found that, somewhat surprisingly,
the mutual inclination of the two long-period planets orbiting 47 UMa
must be less than $45^\circ$ for the system to be stable, but the
results are essentially insensitive to the actual planet masses.

As the above results clearly suggest, dynamical limitations on the
relative inclinations and masses of planetary companions in
extra-solar multiple-planet systems cannot be stated very
precisely. Such estimates suffer from large uncertainties due to
the typically poorly constrained orbital inclinations and planet
masses, and undetermined position angles of the lines of nodes.
More accurate observational data to complement the one-dimensional
radial velocity measurements are needed before the details of the
long-term evolution of multiple-planet systems can be assessed
with a high degree of confidence.

Figure~\ref{fig11} shows the estimated accuracy with which
SIM would measure the coplanarity of the orbits of the present
set of extra-solar planetary systems. After the multiple-planet
fit, the estimated values of $i$ and $\Omega$ for all
components in each system were used to derive their actual
relative inclination (Eq.~\ref{inclrel}). In Figure~\ref{fig11},
for each system, the relative inclination of each pair of planets is
plotted against the (common) inclination of the orbital plane
with respect to the plane of the sky. The corresponding error
bars have been computed by propagating (Eq.~\ref{propaga})
the formal expressions from the covariance matrix of the
multiple-planet fit.

In order to establish coplanarity, we must measure
$i_\mathrm{rel}\simeq 0^\circ$. The first seven panels show how,
as long as all components in a given system have minimum masses
and orbital periods such that the astrometric signal-to-noise
ratio is favorable ($\alpha/\sigma_d\simeq 10$ or greater) for
any inclination of the orbital plane with respect to the plane of
the sky, the quasi-coplanarity of each pair of planetary orbits
can be assessed with high accuracy. Even in the case of two
systems with one of the planets having period exceeding the
timespan of the observations (HD 37124 and 47 UMa), the net result
is that for such well-measured systems the relative inclination can 
be determined to be $i_\mathrm{rel} \leq 3^\circ$, with relative
uncertainties of a few degrees. The next seven panels of
Figure~\ref{fig11} highlight instead how, when dealing with
astrometric signatures approaching the detection limit
($\alpha/\sigma_d\rightarrow 1$), to make accurate coplanarity
measurements will be a significantly more challenging task. In the
specific cases shown, only towards quasi-face-on configurations,
in which the signal from the smaller and/or shorter period planet
becomes large enough (due to significantly larger projected mass),
the results resemble those obtained in the favorable cases. 
Otherwise, the relative inclinations will typically be 
measured to be $i_\mathrm{rel}\sim 
10^\circ$ or greater, with uncertainties of additional tens of
degrees. Finally, as the bottom right panel shows, for the two
outer planets of the $\upsilon$ And system quasi-coplanarity can
be reliably established, in agreement with the findings of
Sozzetti et al.~\citeyearpar{sozzetti01}.

It is worth mentioning another interesting feature arising from
the results presented in Figure~\ref{fig11}. Almost all panels
show a more or less pronounced trend for increasing uncertainties
in the determination of $i_\mathrm{rel}$ as we move away from the 
edge-on configuration, with a somewhat sharp change in behavior as 
$i\rightarrow 0^\circ$. The explanation is to be found in the 
details of the dependence of $i_\mathrm{rel}$ on $i_\mathrm{in}$,
$i_\mathrm{out}$, and the difference
$\Omega_\mathrm{out}-\Omega_\mathrm{in}$. In fact, as $i$
decreases, the uncertainty in the position angle of the line of
nodes grows, because of the increasing difficulty in its correct
identification ($\Omega$ eventually becomes undefined at $i =
0^\circ$). However, the uncertainty in the value of $i_\mathrm{rel}$ 
defined in Eq.~\ref{inclrel} increases correspondingly only up to 
the point in which, for sufficiently low values ($i\leq 5^\circ$) 
of the inclination 
angles, the second term $\sin i_\mathrm{in}\sin i_\mathrm{out} 
\cos(\Omega_\mathrm{out}-\Omega_\mathrm{in})$ of the right-hand 
side becomes essentially negligible with respect to the first term 
$\cos i_\mathrm{in}\cos i_\mathrm{out}$, irrespective of the fact 
that the uncertainty of the line of nodes keeps on growing. 
Thus, ultimately for quasi-face-on configurations an accurate knowledge
of $\Omega_\mathrm{in}$ and $\Omega_\mathrm{out}$ is not required.
In their recent work on extra-solar planetary system detection 
and measurement with GAIA, Sozzetti et al.~\citeyearpar{sozzetti01} do 
not consider orbital configurations of the $\upsilon$ And system with 
inclination angles less than 5 degrees. As a consequence, they come to 
the conclusion that when $i\rightarrow 0^\circ$ accurately estimating
the coplanarity of the outer two planets of the $\upsilon$ And
system would be difficult because of the increasing uncertainties
on the measurement of $\Omega$ for the two planets. Here instead we
have shown how the relative inclination becomes essentially
insensitive to the retrieved values of $\Omega_\mathrm{in}$ and
$\Omega_\mathrm{out}$, as we approach a perfectly face-on configuration.

Finally, simulations of non-coplanar systems with
$\Omega_\mathrm{in}\neq\Omega_\mathrm{out}$ and/or
$i_{\mathrm{in}}\neq i_\mathrm{out}$ yielded similar results. Our
findings help to reaffirm the importance of SIM high-precision
position measurements in verifying the stability of multiple-planet
system. Accurately determining the inclination angles
and lines of nodes of multiple planetary orbits
will allow in turn to derive meaningful mass and relative inclination
angle estimates, which will be used to better constrain the results
from theoretical studies on the long-term evolution of planetary
systems.

\subsubsection{Combining Radial Velocity and Astrometric Data}
\label{constrain}

In recent works, Eisner \& Kulkarni~\citeyearpar[2001a, ][2002]{eisner01b} 
have utilized a semi-analytical method to show how planet detection
efficiency would benefit from the simultaneous availability of
both astrometric {\it and} radial velocity measurements,
especially as far as long-period, edge-on orbits are concerned. This
combined approach has indeed proved very successful when dealing
with real data. For example, utilizing both astrometric
measurements from the Fine Guidance Sensor 3 on board the {\it
Hubble Space Telescope} and ground-based precision radial velocity
data, Benedict et al.~\citeyearpar{benedict01} 
have improved the accuracy on the
mass estimates for the M dwarf binary Wolf 1062 by a factor 4 with
respect to the same values obtained by Franz et al.~\citeyearpar{franz98} 
using only {\it HST} astrometry. Very recently, the same combined method
revealed itself an essential ingredient in the spectacular
determination of the first actual mass of an extra-solar planet,
the outer component in the Gliese 876 system~\citep{benedict02}.

Usually, a combined astrometric + spectroscopic orbit is obtained
by modeling time-series of data from both techniques in a
simultaneous least-squares solution, with a few additional constraints, 
the most important of which is the identity~\citep{pourbaix00}:
\begin{equation}\label{constraint}
\frac{\alpha\sin i}{\pi_\star} =
\frac{T K_1\sqrt{1 - e^2}}{2\pi\times 4.7405},
\end{equation}
which relates quantities only derived from astrometry (inclination
angle, astrometric signature, and stellar parallax on the left-hand side)
to quantities derivable from both techniques or radial velocity alone
(orbital period and eccentricity, and radial velocity semi-amplitude of 
the primary on right-hand side).

In our exploratory work, we have adopted a more limited approach.
For each of the 11 planetary systems under study, we have utilized
the orbital elements derived from spectroscopy ($T$, $e$, $\tau$,
and $\omega$) as constraints in the sense that we have kept them
fixed to their input values in the global least-squares iterative
solution, and solved only for $\alpha$, $i$, $\Omega$.
Figure~\ref{fig12} shows the estimated improvement that may be
obtained in the determination of planet mass, semi-major axis, and
inclination when SIM relative astrometry of the presently known
planetary systems is ``combined'' (in the abovementioned sense)
with spectroscopy, as opposed to the scenario in which all orbital
elements are solved for in the least-squares minimization
procedure. The three contour plots identify, in the plane defined
by $\alpha/\sigma_d$ and $T$, regions of increasingly larger
values of the ratio $\sigma_G/\sigma_C$ between the estimated
fractional uncertainties on $M_\mathrm{p}$, $a$, and $i$ in the
case of `global' 
orbital fits in which all parameters were adjusted and the
same quantities as computed after the `constrained' fits. The
results are averaged over the inclination angle.

The general indication is that for well-sampled orbits, with large
astrometric signatures the constrained fit does not improve significantly
the final accuracy of the results. In such cases, typically
$1\leq\sigma_G/\sigma_C\leq 2$. The ability to recover accurate
values of $M_\mathrm{p}$, $a$, and $i$ will be improved especially
when the period is larger than the timespan of the observations
(and the astrometric signal-to-noise ratio is very large), with
typical values of $\sigma_G/\sigma_C\geq 6$ for $a$ and $M_\mathrm{p}$,
while the inclination is less sensitive to the constrained solution.
For low values of $\alpha/\sigma_d$ and short-period orbits the
improvement on the fractional accuracy for the fitted parameters
will be marginal.

In their work on multiple-planet system detection and measurement
with GAIA, Sozzetti et al.~\citeyearpar{sozzetti01} 
suggested that, for systems with
configurations very close to face-on, the accuracy of the
inclination measurement will be substantially increased by
combining radial velocity and astrometric data. On the other hand,
although in their work they focused mainly on combining astrometry
and radial velocity in the case of edge-on (single-planet) orbits,
Eisner and Kulkarni~\citeyearpar{eisner02} 
argue that the short-period accuracy of
astrometry + radial velocity should be approximately independent
of orbital inclination. In Figure~\ref{fig13} and~\ref{fig14} we
show the behavior of the ratios $\sigma_G/\sigma_C$ for
$M_\mathrm{p}$ (upper panels), $a$ (central panels), and $i$
(lower panels) as a function of the common inclination of the
orbital plane for four representative systems (55 Cnc, 47 UMa, HD
12661, and HD 38529), whose characteristics allow us to provide
quantitative answers to the above arguments. Although the
predictions made by Sozzetti et al.~\citeyearpar{sozzetti01} and 
Eisner \& Kulkarni~\citeyearpar{eisner02} 
are in general confirmed, when it comes to multiple-planet
orbital fits some details become more subtle and complex. In fact,
not only the average value of the ratios $\sigma_G/\sigma_C$ (as
shown in Figure~\ref{fig12}), but also their behavior as a
function of the orbital inclination depends on the ranges of
periods and astrometric signatures we are dealing with. In the
case of 55 Cnc d, for example, with a period $T \simeq 3 L$, the
accuracy on all fitted parameters benefits greatly from a
constrained fit, which minimizes the residual covariance between
the orbital solution and the solution for the parallax and the
proper motion of the primary. The improvement in orbit
reconstruction reaches its maximum for an almost face-on
configuration. On the other hand, for 47 UMa c and HD 38529 c,
with $T\simeq 1.4 L$ and $T\simeq 1.2 L$ respectively,
$\sigma_G/\sigma_C$ increases as $i\rightarrow 0^\circ$ only in
the case of $i$ itself, while the semi-major axis is almost
insensitive to the inclination angle. In the range of periods $T <
L$, again different parameters behave slightly differently, in
these cases primarily due to the magnitude of $\alpha/\sigma_d$.
While the accuracy on orbital inclination is only marginally
improved towards face-on configurations, the estimated semi-major
axis (and in turn the derived planet mass) is more accurately
retrieved in a constrained fit on an edge-on configuration, in
which case astrometry loses completely the second dimension of the
measurements. \footnote{A noticeable exception is constituted by
47 UMa b, that exhibits significant improvement in the accuracy
with which its semi-major axis can be determined when the
configuration is face-on. This behavior is due to the particular
orbital arrangement of the system, with the longer period planet
producing the smaller astrometric signal, under the assumption of
coplanar orbits} For an accurate determination of the semi-major
axis, the combination between astrometric and spectroscopic data
will be effective particularly when $\alpha/\sigma_d\rightarrow
1$, as in the case of HD 38529 b or 55 Cnc c, while for example
the results for the two planets in the HD 12661 system, both with large 
signatures, are only marginally improved by the
constrained fit.

\section{Summary and Conclusions}

The properties of extra-solar giant planets detected by radial velocity
surveys of nearby solar-type stars~\citep{marcy02a} 
seem to indicate that the Solar
System configuration is just one of the many possible outcomes of
disk evolution around young stars. Indeed, the strong
coupling between the early evolutionary processes of a star and
its disk can have a significant impact on planet formation
time-scales and the final orbital configurations after the
disk dissipates.

Over the next decade of so, a series of new instruments will come on
line,
which will provide data of great value to shed new light in the complex
scenarios of the formation and evolution of planetary systems.
Among indirect detection techniques,
ground-based precision spectroscopy, the most successful technique
so far, will be complemented by high-precision ground-based (Keck 
Interferometer, VLTI) and space-borne (SIM, GAIA) astrometry, 
and transit photometry from ground (e.g., OGLE III, STARE, STEPSS, 
Vulcan Camera Project) and in space (Corot, Kepler, Eddington). 
In the field of direct detection techniques, near- and far-infra-red
diffraction-limited ground- and space-based imaging (ALMA, SIRTF, JWST)
will pave the way to ambitious projects of coronagraphic/interferometric
imaging from space (TPF, Darwin), with the long-term goal of directly
imaging terrestrial planets in the Habitable Zone of nearby stars.

In this paper we have completed the analysis begun in our
previous works (Casertano \& Sozzetti 1999; S02), in order to 
connect and relate the basic 
capabilities of the Space Interferometry Mission (SIM) to the
properties of extra-solar planetary systems. We have utilized
detailed end-to-end numerical simulations of sample SIM
narrow-angle astrometric observing campaigns
(Section~\ref{scenario}), an improved methodology for planet
detection which combines $\chi^2$ statistics, periodogram
analysis, and Fourier series expansions (Sections~\ref{stattools}),
an upgraded analytical model that allows for multiple-planet
orbital fits (Sections~\ref{multifits}), and the set of presently
known extra-solar planetary systems as templates. The experiments
described in this paper have allowed us to quantify the limiting
capability of SIM to discover systems of planets 
around solar-type stars in the solar neighborhood 
(Sections~\ref{probdet},~\ref{periods}, and~\ref{fourier}), measure 
their orbital properties and masses (Sections~\ref{reconstruct} 
and~\ref{constrain}), and accurately determine 
the coplanarity of multiple-planet orbits (Section~\ref{coplan}).
Our main results can be summarized as follows.

\begin{itemize}
\item[1.] Additional planets in systems have little impact
on SIM ability to detect each component in a system, in comparison to 
the single-planet configurations. The inaccurate fit and subtraction 
of orbits with astrometric signal-to-noise ratio 
$\alpha/\sigma_d\rightarrow 1$ can on the other hand 
increase the false detection rate by up to 
a factor 2. The periodogram analysis adds robustness to the
detection method when $T\leq L$, by singling out periodicities
even in the case of $\alpha/\sigma_d$ close to the $\chi^2$ 
detection limit. For very
short-period orbits, a more dense time-series of observations will
be the obvious choice in order to overcome poor sampling. When
$T\geq L$, the least squares technique combined with Fourier
analysis can correctly identify periods as long as three times the
timespan of the observations. This approach is arguably preferable, 
as opposed to the periodogram method, 
which needs modifications in the long-period regime.
\item[2.] Accurate measurements of multiple-planet
orbits and determination of planet masses are only moderately 
affected by the presence of more than one object in a system,
with typical degradation of 30-40\% with respect to single-planet
solutions. When $T\leq L$, it is possible to determine masses
and orbital inclinations to better than 10\% for systems with 
planets having periods as 
short as 0.1 yr, and for systems with components producing
astrometric  signals as low as $\sim 5\,\sigma_d$, while
$\alpha/\sigma_d\simeq 100$ is required in order to measure with
similar accuracy objects with periods as long as three times the mission
duration. Orbital eccentricity typically requires larger signals for
the same accuracy level, and its correct identification can become a
non trivial task when $T\geq L$. The accuracy on estimated orbital 
elements improves significantly as we move towards face-on orbits, 
except for the inclination angle. 
\item[3.] Accurate coplanarity measurements are possible 
for systems with all components producing $\alpha/\sigma_d\simeq
10$ or greater. In the case of truly coplanar systems, 
the relative inclination between pairs of planetary orbits is 
measured to be $i_\mathrm{rel}\leq 3^\circ$, with uncertainties of a 
few degrees, for periods $0.1\leq T\leq 15$ yr. In systems where at
least one component has $\alpha/\sigma_d\rightarrow 1$,
uncertainties on $i_\mathrm{rel}$ of order of $30^\circ-40^\circ$,
or larger, are likely to preclude a robust assessment of the system 
coplanarity. 
\item[4.] Whenever feasible, an approach that combines astrometry
and radial velocity will yield significantly more accurate
estimates of planet masses and orbital elements. The uncertainties
on orbital elements and masses can be reduced by up to an order
of magnitude, especially in the case of long-period orbits in
face-on configurations, and for low amplitude orbits seen edge-on.
Well-sampled, well-measured orbits ($T\leq L$, $\alpha/\sigma_d\gg
1$) are only marginally affected by the combination of
astrometric and radial velocity measurements.
\end{itemize}

Our results reaffirm the important role future high-precision
space-borne astrometric missions promise to play in the realm of
extra-solar planets. With its unprecedented astrometric precision,
SIM will not only complement other on-going and planned
spectroscopic, astrometric, and photometric surveys, but its
position measurements will have a unique impact in the study of
some important aspects of multiple-planet systems. By directly
measuring the line of nodes and inclination angle for each
component in a system, SIM will determine whether planetary orbits
are coplanar with uncertainties of a few degrees. For instance,
this will provide theory with the observational evidence needed to
address the long-term evolution issue, and draw sensible conclusions 
on the chaotic and stable/unstable dynamical behavior of multiple-planet 
orbits. The same SIM measurements will also be instrumental to help 
confirm or rule out one of paradigms which form the basis for present-day
theoretical models, i.e. that the sole conceivable environment for
planet formation are flattened circumstellar disks around young
pre-main sequence stars. If multiple-planet orbits are found to be
coplanar, this will indicate that planetary systems indeed originate 
and evolve in a way similar to our own~\citep{lissauer93,
pollack96}. If large relative orbital inclinations
were found, this would provide evidence that other systems are
truly different from ours, and thus their present configurations
should be explained in terms of either an early, chaotic phase of
orbital evolution or formation by another mechanism such as disk
instability (\citealp[2001]{boss00};~\citeauthor{mayer02}
\citeyear{mayer02}).

\acknowledgements

This project was partially funded by the Italian Space Agency 
under contract ASI-I/R/117/01 (to A. S. and M. G. L.). We thank an 
anonymous referee for a very careful reading of the paper and for 
illuminating comments that helped improve the original manuscript.
A. S. is greatly indebted to the Smithsonian Astrophysical Observatory 
and its Computation Facility for kind hospitality and technical 
support during the completion of this work. 
This research has made use of the SIMBAD database, 
operated at CDS, Strasbourg, France.

\clearpage

\figcaption[]{Probability of detection as a function of 
orbital inclination $i$, for each planet in each of the 11 
multiple-planet systems considered in our study. The 
three planets HD 38529 b (dashed-dotted-dotted line), $\upsilon$ And b 
(long-dashed line), and 55 Cnc c (dashed-dotted line) are athe 
most difficult to detect. After the not accurate subtraction 
of their low-amplitude astrometric signatures, the corresponding 
false detection rates are up to a factor 2 larger than the 
expected 5\% \label{fig1}}

\figcaption[]{Periodogram analysis for the HD 38529 planetary 
system. In the upper left panel we show the periodicity search 
after the single-star fit: a long-period signal is clearly 
found with high significance, but due to the incomplete orbital 
sampling, its value is about a factor 2 shorter than the actual period 
of HD 38529 c. The upper right panel highlights how, after 
the removal of the signal with the larger amplitude, the period 
of HD 38529 b is still accurately identified, in spite of the 
very low astrometric signature induced by the inner planet of 
the system on its central star. The lower left panel 
present the results of the periodogram analysis after the 
dual Keplerian fit to the observations: no significant peaks 
are found \label{fig2}}

\figcaption[]{Same as Figure~\ref{fig2}, but for the $\upsilon$ And 
planetary system. The periodogram analysis correctly identifies 
the periods of the two outer planets (two upper panels), and rules out the 
presence of additional planets after subtraction of all 
planetary signatures (lower right panel). Due to very short 
period of the innermost planet, instead, after removal 
of the signatures of $\upsilon$ And c and $\upsilon$ And d 
the periodogram search identifies a significant peak in the spectrum 
which does not correspond to the true period of $\upsilon$ And b 
(lower left panel) \label{fig3}}

\figcaption[]{Same as Figure~\ref{fig2}, but for the `easy' 
case of the HD 82943 planetary system \label{fig4}}

\figcaption[]{Same as Figure~\ref{fig2}, but for the case 
of the 55 Cnc planetary system. The planet inducing the larger 
signature, 55 Cnc d, is identified immediately in the power 
spectrum (upper left panel), but, similarly to the case of 
HD 38529 c shown in Figure~\ref{fig2}, the incomplete 
orbital sampling causes the periodogram analysis to identify 
an incorrect value for the orbital period of 55 Cnc d \label{fig5}}

\figcaption[]{Reduced chi-square $\chi^2_\nu$ as a function of 
a dense grid of trial periods for the five planetary systems 
harboring one planet with a period exceeding the timespan of 
SIM observations (two upper, two middle, and lower left panels). 
For comparison, the lower right panel shows the results of 
the same procedure applied to our own solar system, placed 
at increasing distance from the observer (details in the text)
\label{fig6}}

\figcaption[]{Fractional errors (\%) for $M_\mathrm{p}$, $a$, $T$, 
$e$, and $i$, as a function of the inclination of the orbital plane, 
in the case of the 55 Cnc planetary system. In each panel, the results 
for a given parameter are shown for all the system components, 
for the cases of a single-planet solution (dashed-dotted, 
dotted, and short-dashed lines for 55 Cnc b, 55 Cnc c, and 55 Cnc d, 
respectively) and a complete three-planet model (solid, long-dashed, and 
dashed-dotted-dotted lines for 55 Cnc b, 55 Cnc c, and 55 Cnc d, 
respectively) 
  \label{fig7}}

\newpage

\figcaption[]{Same as Figure~\ref{fig7}, but for the case of the 
HD 12661 planetary system. The dashed-dotted and dashed-dotted-dotted 
lines are relative to the single-planet solution for 
HD 12661 b and HD 12661 c, respectively. The solid and long-dashed 
lines are relative to the dual Keplerian fit for 
HD 12661 b and HD 12661 c, respectively    \label{fig8}}

\figcaption[]{Iso-accuracy contours in the $e-T$ plane, 
for the most relevant orbital parameters and planet mass. The 
results in each panel have been obtained utilizing the 11 planetary 
systems known to-date. The contour regions are 
color-coded by the accuracy achieved in the determination of 
the given parameter, expressed as a fraction (\%) of its true value 
\label{fig9}}

\figcaption[]{Same as Figure~\ref{fig9}, but with the contour 
regions identified in the $\alpha/\sigma_d - T$ plane  \label{fig10}}

\figcaption[]{Relative inclination $i_\mathrm{rel}$ 
between pairs of planetary orbits as a 
function of the common inclination angle with respect to the 
line of sight, for the entire set of presently known multiple-planet 
systems. In each panel, the corresponding uncertainties are 
computed utilizing the formal expressions from the covariance 
matrix of the multiple Keplerian fit
 \label{fig11}}

\figcaption[]{Contours identifying regions, in the $\alpha/\sigma_d-T$ 
plane, with equal values of the ratios $\sigma_G/\sigma_C$ of 
the estimated fractional uncertainties on $M_\mathrm{p}$, $a$, 
and $i$ in the case of `global' fits in which all parameters were 
adjusted to the same quantities as computed after the `constrained' 
fits (as discussed in the text), for the set of presently 
known planetary sytems. The results are averaged over the inclination 
angle (coplanar orbits are assumed). The contour regions are color-coded 
by the value of the ratio $\sigma_G/\sigma_C$ for each parameter
 \label{fig12}}

\figcaption[]{The behavior of the ratios $\sigma_G/\sigma_C$ for 
$M_\mathrm{p}$ (upper panels), $a$ (middle panels), and $i$ (lower panels), 
expressed as a function of the common inclination angle, 
for the two representative multiple-planet systems 55 Cnc and 47 UMa
 \label{fig13}}

\figcaption[]{Same as Figure~\ref{fig13}, but for the HD 12661 and 
HD 38529 planetary systems \label{fig14}}


\clearpage

\begin{deluxetable}{cccccc}

\tablecaption{Fundamental stellar characteristics, astrometric parameters
and orbital
elements of the presently known set of extra-solar planetary systems
utilized in the simulations. Planet masses and
astrometric signatures are computed as lower limits corresponding
to edge-on configurations ($\sin i = 1$).
The data have been taken utilizing various web resources containing 
up-to-date values of all parameters,
as explained in the notes\label{planparam}}
\tablewidth{0pt}
\tablehead{\colhead{Stellar Parameters\tablenotemark{a}}&
\colhead{Central Star}&\colhead{ Orbital elements\tablenotemark{b}}&
\colhead{Planet b}& \colhead{Planet c}&
 \colhead{Planet d}}
\startdata
\\[-10pt]
&&System: $\upsilon$ And &&&\\
\\[-10pt]
$\lambda$ (deg)&38.36 & $\alpha$ ($\mu$as)&2.32&89.58&541.8 \\
$\beta$ (deg)&29.25&$a$ (AU) &0.059&0.83&2.53 \\ $\mu_\lambda$
(mas/yr)& $-$313.16& $T$ (d) &4.6&241.5&1284.0\\ $\mu_\beta$
(mas/yr)&$-$277.30 &  $e$ &0.012&0.28&0.27\\ $\pi$ (mas)&74.25&
$\tau$ (JD)&2450002.1&2450160.5&2450064.0\\ $V$ Magnitude&4.09&
$\omega$ (deg)&73.0&250.0&260.0\\ Spectral Type& F8V &
$M_\mathrm{p}\sin i$ ($M_\mathrm{J}$)&0.69&1.89&3.75\\ $d$
(pc)&13.47 &$\Omega$ (deg)& $[0,\pi]$&$[0,\pi]$&$[0,\pi]$\\
$M_\star$ ($M_\odot$)&1.3& $i$
(deg)&$[0,\pi/2]$&$[0,\pi/2]$&$[0,\pi/2]$ \\
\\[-10pt]
\tableline
\tableline
\\[-10pt]
&&System: 55 Cnc &&&\\
\\[-10pt]
$\lambda$ (deg)&127.79 & $\alpha$ ($\mu$as)&7.75&4.04&1916.6 \\
$\beta$ (deg)&10.70&$a$ (AU) &0.115&0.24&5.9 \\ $\mu_\lambda$
(mas/yr)& $-$530.91& $T$ (d) &14.653&44.28&5360.0\\ $\mu_\beta$
(mas/yr)&$-$93.54 &  $e$ &0.02&0.34&0.16\\ $\pi$ (mas)&79.81&
$\tau$ (JD)&2450001.479&2450031.4&2452785.0\\ $V$ Magnitude&5.95&
$\omega$ (deg)&99.0&61.0&201.0\\ Spectral Type& G8V &
$M_\mathrm{p}\sin i$ ($M_\mathrm{J}$)&0.84&0.21&4.05\\ $d$
(pc)&12.53 &$\Omega$ (deg)& $[0,\pi]$&$[0,\pi]$&$[0,\pi]$\\
$M_\star$ ($M_\odot$)&0.95& $i$
(deg)&$[0,\pi/2]$&$[0,\pi/2]$&$[0,\pi/2]$ \\
\\[-10pt]
\tableline
\tableline
\\[-10pt]
&&System: HD 38529 &&&\\
\\[-10pt]
$\lambda$ (deg)&86.23 & $\alpha$ ($\mu$as)&1.71&805.7&\nodata \\
$\beta$ (deg)&$-$21.78&$a$ (AU) &0.129&3.71& \nodata\\
$\mu_\lambda$ (mas/yr)& $-$83.66& $T$ (d) &14.31&2207.4&\nodata\\
$\mu_\beta$ (mas/yr)&$-$139.69 &  $e$ &0.28&0.33&\nodata\\ $\pi$
(mas)&23.58& $\tau$ (JD)&24510005.8&24510043.7&\nodata\\ $V$
Magnitude&5.95& $\omega$ (deg)&90.0&13.0&\nodata\\ Spectral Type&
G4IV & $M_\mathrm{p}\sin i$ ($M_\mathrm{J}$)&0.78&12.8&\nodata\\
$d$ (pc)&42.40 &$\Omega$ (deg)& $[0,\pi]$&$[0,\pi]$&\nodata\\
$M_\star$ ($M_\odot$)&1.39& $i$
(deg)&$[0,\pi/2]$&$[0,\pi/2]$&\nodata \\
\\[-10pt]
\tableline
\tableline
\\[-10pt]
&&System: Gliese 876 &&&\\
\\[-10pt]
$\lambda$ (deg)&339.21 & $\alpha$ ($\mu$as)&264.4&48.5&\nodata \\
$\beta$ (deg)&$-$6.77&$a$ (AU) &0.21&0.13&\nodata \\ $\mu_\lambda$
(mas/yr)& 634.86& $T$ (d) &61.02&30.12&\nodata\\ $\mu_\beta$
(mas/yr)&$-$987.72 &  $e$ &0.10&0.27&\nodata\\ $\pi$ (mas)&213.22&
$\tau$ (JD)&2450106.2&2450031.4&\nodata\\ $V$ Magnitude&10.15&
$\omega$ (deg)&333.0&330.0&\nodata\\ Spectral Type& M4 &
$M_\mathrm{p}\sin i$ ($M_\mathrm{J}$)&1.89&0.56&\nodata\\ $d$
(pc)&4.69 &$\Omega$ (deg)& $[0,\pi]$&$[0,\pi]$&\nodata\\ $M_\star$
($M_\odot$)&0.32& $i$ (deg)&$[0,\pi/2]$&$[0,\pi/2]$&\nodata \\
\\[-10pt]
\tableline
\tableline
\\[-10pt]
&&System: HD 168443 &&&\\
\\[-10pt]
$\lambda$ (deg)&225.07 & $\alpha$ ($\mu$as)&58.64&1269.5&\nodata
\\ $\beta$ (deg)&13.33&$a$ (AU) &0.295&2.87&\nodata \\
$\mu_\lambda$ (mas/yr)& $-$99.95& $T$ (d) &58.10&1771.5&\nodata\\
$\mu_\beta$ (mas/yr)&$-$220.79 &  $e$ &0.53&0.20&\nodata\\ $\pi$
(mas)&25.97& $\tau$ (JD)&2450047.6&2450250.0&\nodata\\ $V$
Magnitude&6.91& $\omega$ (deg)&172.9&289.0&\nodata\\ Spectral
Type& G5IV & $M_\mathrm{p}\sin i$
($M_\mathrm{J}$)&7.73&17.2&\nodata\\ $d$ (pc)&38.5 &$\Omega$
(deg)& $[0,\pi]$&$[0,\pi]$&\nodata\\ $M_\star$ ($M_\odot$)&1.01&
$i$ (deg)&$[0,\pi/2]$&$[0,\pi/2]$&\nodata \\
\\[-10pt]
\tableline
\tableline
\\[-10pt]
&&System: HD 12661 &&&\\
\\[-10pt]
$\lambda$ (deg)&37.70 & $\alpha$ ($\mu$as)&47.43&100.4&\nodata \\
$\beta$ (deg)&12.26&$a$ (AU) &0.82&2.56&\nodata \\ $\mu_\lambda$
(mas/yr)& $-$161.28& $T$ (d) &263.3&1444.5&\nodata\\ $\mu_\beta$
(mas/yr)&$-$127.78 &  $e$ &0.35&0.20&\nodata\\ $\pi$ (mas)&26.91&
$\tau$ (JD)&2459943.7&2459673.9&\nodata\\ $V$ Magnitude&7.43&
$\omega$ (deg)&292.6&147.0&\nodata\\ Spectral Type& G6V &
$M_\mathrm{p}\sin i$ ($M_\mathrm{J}$)&2.30&1.56&\nodata\\ $d$
(pc)&37.16 &$\Omega$ (deg)& $[0,\pi]$&$[0,\pi]$&\nodata\\
$M_\star$ ($M_\odot$)&1.07& $i$
(deg)&$[0,\pi/2]$&$[0,\pi/2]$&\nodata \\
\\[-10pt]
\tableline
\tableline
\\[-10pt]
&&System: HD 160691\tablenotemark{c} &&&\\
\\[-10pt]
$\lambda$ (deg)&267.20 & $\alpha$ ($\mu$as)&155.8&139.2&\nodata \\
$\beta$ (deg)&-28.87&$a$ (AU) &1.48&2.30&\nodata \\ $\mu_\lambda$
(mas/yr)& $-$20.96& $T$ (d) &637.3&1300.0&\nodata\\ $\mu_\beta$
(mas/yr)&$-$190.61 &  $e$ &0.31&0.8&\nodata\\ $\pi$ (mas)&65.36&
$\tau$ (JD)&2450959.0&2451613.0&\nodata\\ $V$ Magnitude&5.15&
$\omega$ (deg)&320.0&99.0&\nodata\\ Spectral Type& G3IV/V &
$M_\mathrm{p}\sin i$ ($M_\mathrm{J}$)&1.74&1.0&\nodata\\ $d$
(pc)&15.30 &$\Omega$ (deg)& $[0,\pi]$&$[0,\pi]$&\nodata\\
$M_\star$ ($M_\odot$)&1.08& $i$
(deg)&$[0,\pi/2]$&$[0,\pi/2]$&\nodata \\
\\[-10pt]
\tableline
\tableline
\\[-10pt]
&&System: 47 UMa &&&\\
\\[-10pt]
$\lambda$ (deg)&149.30 & $\alpha$ ($\mu$as)&366.1&195.5&\nodata \\
$\beta$ (deg)&31.28&$a$ (AU) &2.09&3.73&\nodata \\ $\mu_\lambda$
(mas/yr)& $-$259.16& $T$ (d) &1089.0&2594.0&\nodata\\ $\mu_\beta$
(mas/yr)&188.90 &  $e$ &0.06&0.10&\nodata\\ $\pi$ (mas)&71.02&
$\tau$ (JD)&2453622.0&2451363.5&\nodata\\ $V$ Magnitude&5.03&
$\omega$ (deg)&172.0&127.0&\nodata\\ Spectral Type& G0V &
$M_\mathrm{p}\sin i$ ($M_\mathrm{J}$)&2.54&0.76&\nodata\\ $d$
(pc)&14.08 &$\Omega$ (deg)& $[0,\pi]$&$[0,\pi]$&\nodata\\
$M_\star$ ($M_\odot$)&1.03& $i$
(deg)&$[0,\pi/2]$&$[0,\pi/2]$&\nodata \\
\\[-10pt]
\tableline
\tableline
\\[-10pt]
&&System: HD 82943\tablenotemark{d} &&&\\
\\[-10pt]
$\lambda$ (deg)&150.22 & $\alpha$ ($\mu$as)&65.58&22.28&\nodata \\
$\beta$ (deg)&$-$24.79&$a$ (AU) &1.16&0.73&\nodata \\
$\mu_\lambda$ (mas/yr)& $-$58.14& $T$ (d) &444.6&221.6&\nodata\\
$\mu_\beta$ (mas/yr)&$-$164.07 &  $e$ &0.41&0.54&\nodata\\ $\pi$
(mas)&36.42& $\tau$ (JD)&2451620.3&2451630.9&\nodata\\ $V$
Magnitude&6.54& $\omega$ (deg)&96.0&138.0&\nodata\\ Spectral Type&
G0 & $M_\mathrm{p}\sin i$ ($M_\mathrm{J}$)&1.63&0.88&\nodata\\ $d$
(pc)&27.46 &$\Omega$ (deg)& $[0,\pi]$&$[0,\pi]$&\nodata\\
$M_\star$ ($M_\odot$)&1.05& $i$
(deg)&$[0,\pi/2]$&$[0,\pi/2]$&\nodata \\
\\[-10pt]
\tableline
\tableline
\\[-10pt]
&&System: HD 74156\tablenotemark{d} &&&\\
\\[-10pt]
$\lambda$ (deg)&131.71 & $\alpha$ ($\mu$as)&6.35&383.9&\nodata \\
$\beta$ (deg)&$-$12.86&$a$ (AU) &0.276&3.47&\nodata \\
$\mu_\lambda$ (mas/yr)& $-$28.19& $T$ (d) &51.61&2300.0&\nodata\\
$\mu_\beta$ (mas/yr)&$-$200.05 &  $e$ &0.65&0.40&\nodata\\ $\pi$
(mas)&15.49& $\tau$ (JD)&2451981.4&2450849.0&\nodata\\ $V$
Magnitude&7.62& $\omega$ (deg)&183.7&240.0&\nodata\\ Spectral
Type& G0 & $M_\mathrm{p}\sin i$
($M_\mathrm{J}$)&1.56&7.5&\nodata\\ $d$ (pc)&64.56 &$\Omega$
(deg)& $[0,\pi]$&$[0,\pi]$&\nodata\\ $M_\star$ ($M_\odot$)&1.05&
$i$ (deg)&$[0,\pi/2]$&$[0,\pi/2]$&\nodata \\
\\[-10pt]
\tableline
\tableline
\\[-10pt]
&&&&&\\
&&System: HD 37124 &&&\\
\\[-10pt]
$\lambda$ (deg)&84.62 & $\alpha$ ($\mu$as)&15.37&98.62&\nodata \\
$\beta$ (deg)&$-$2.17&$a$ (AU) &0.54&2.95&\nodata \\ $\mu_\lambda$
(mas/yr)& $-$96.14& $T$ (d) &153.0&1942.0&\nodata\\ $\mu_\beta$
(mas/yr)&$-$416.51 &  $e$ &0.10&0.40&\nodata\\ $\pi$ (mas)&30.12&
$\tau$ (JD)&2451227.0&2451828.0&\nodata\\ $V$ Magnitude&7.68&
$\omega$ (deg)&97.0&265.0&\nodata\\ Spectral Type& G4V &
$M_\mathrm{p}\sin i$ ($M_\mathrm{J}$)&0.86&1.01&\nodata\\ $d$
(pc)&33.20 &$\Omega$ (deg)& $[0,\pi]$&$[0,\pi]$&\nodata\\
$M_\star$ ($M_\odot$)&0.91& $i$
(deg)&$[0,\pi/2]$&$[0,\pi/2]$&\nodata \\
\enddata
\tablenotetext{a}{Data from \url{http://simbad.u-strasbg.fr/}. Positions and
proper motion components, originally given in equatorial coordinates, have
been projected to the ecliptic reference frame
~\citep[see for example ][]{green85}}
\tablenotetext{b}{Data from \url{http://www.obspm.fr/encycl/encycl.html} and
\url{http://exoplanets.org/} as of September 2002}
\tablenotetext{c}{The existence of the outer planet around HD 160691 is 
still doubtful. Its presence has been inferred by Jones et al. (2002), 
due to a clear trend in the radial velocity residuals, but the orbit still 
awaits phase closure}
\tablenotetext{d}{The discovery of the planetary systems around HD 74156 and
HD 82943 has not yet been reported in refereed Journals}
\end{deluxetable}
\clearpage

\begin{deluxetable}{lccccc}

\tablecaption{Results from the periodogram analysis of the 11
planetary systems known to date, in the case of an average value
of the orbital inclination angle ($i = 45^\circ$). The columns
display, respectively, the name of the system, the type of fit
performed (single-star, single- or multiple-planet), the value of
the reduced chi-square $\chi^2_\nu$ of the fit, the highest peak
$z_p$ in the periodogram power spectrum, the value of the orbital
period $T(z_p)$ corresponding to the peak in the spectrum, and the
false-alarm probability $P(z_p)$ of the peak \label{scargle}}
\tablewidth{0pt} \tablehead{\colhead{System}& \colhead{Type of
Fit}&\colhead{$\chi^2_\nu$}&\colhead{$z_p$}& \colhead{$T(z_p)$}&
\colhead{$P(z_p)$}} \startdata $\upsilon$ And& Single Star
&3096.55 &50.6828&3.75855 &0.0  \\ & 1 Planet &1736.20
&48.5907&0.671169 &0.0 \\ &  2 Planets &1.87563&14.6750 &0.119567
&$6.35\times 10^{-5}$ \\ &  3 Planets &1.14677&8.41723 &0.220069
&0.0326138 \\ &&&&& \\ 55 Cnc& Single Star &1856.12&65.3129
&5.87275 &0.0  \\ & 1 Planet &25.3619&51.3840 &0.0401112 &0.0 \\ &
2 Planets &4.13032&39.1092 &0.121448 &0.0 \\ &  3 Planets
&0.791748&4.85446 &0.0705180 &0.690751 \\ &&&&& \\ HD 38529&
Single Star &1063.33 &35.7711&3.49768 &0.0  \\ & 1 Planet
&1.46852&28.3746 &0.0389847 &$1.37\times 10^{-9}$ \\ &  2 Planets
&0.974422&6.45905 &0.0402565 &0.209530 \\ &&&&& \\ Gliese 876&
Single Star &583.995 &57.6141&0.166749 &0.0  \\ & 1 Planet
&392.007 &53.6403&0.0821299 &0.0 \\ &  2 Planets &0.720508&7.41911
&0.393050 &0.0860501 \\ &&&&& \\ HD 168443& Single Star &13919.0
&54.8187&4.50731 &0.0  \\ & 1 Planet &534.999&42.8020 &0.157193
&0.0 \\ &  2 Planets &0.943551&6.85225 &0.109261 &0.146701 \\
&&&&& \\ HD 12661& Single Star &67.2261 &28.1816&3.30105
&$1.23\times 10^{-7}$  \\ & 1 Planet &481.436&44.0427 &0.741820
&0.0 \\ &  2 Planets &1.13029 &6.83758&0.0402566 &0.148699 \\
&&&&& \\ HD 160691& Single Star &225.807 &50.5789&1.65052 &0.0  \\
& 1 Planet &1650.87 &56.9019&3.27908 &0.0 \\ &  2 Planets
&0.857078&7.96351 &0.397465 &0.0508595 \\ 
47 UMa& Single Star &688.689 &42.7118&2.81892 &0.0  \\ & 1 Planet
&402.954 &51.3840&4.95227 &$1.35\times 10^{-9}$ \\ &  2 Planets
&1.04254 &4.81968&0.1009260 &0.703383 \\ &&&&& \\ HD 82943& Single
Star &29.2090 &53.8883&1.23789 &0.0  \\ & 1 Planet &36.8078
&43.9404&0.618946 &0.0 \\ &  2 Planets &1.01055 &5.77669&0.0398152
&0.372220 \\ &&&&& \\ HD 74156& Single Star &224.126
&48.2071&3.30105 &0.0  \\ & 1 Planet &4.25179 &36.1520&0.140470
&0.0 \\ &  2 Planets &0.994640 &5.25626&0.0330859 &0.543543 \\
&&&&& \\ HD 37124& Single Star &82.4841 &57.8993&4.36463 &0.0  \\
& 1 Planet &36.8113 &39.9654&0.415679 &0.0 \\ &  2 Planets
&0.931258 &5.95232&0.0394809 &0.321418
\enddata
\end{deluxetable}
\clearpage

\begin{deluxetable}{lccccc}

\tablecaption{Ranges of plausible starting guesses for the 
orbital period of the 5 planets with $T > 5$ yr in the presently known 
extrasolar multiple-planet systems. For reference, the same results 
are reported for Jupiter in simulations of the solar system as `seen' 
by SIM at increasing distances (in pc) from the observer. The results are 
derived in the context of the Fourier 
analysis approach discussed in the text. Preliminary values of $T$ 
are considered acceptable if the corresponding $\chi^2_\nu$ does not 
differ from the minimum $\chi^2_\mathrm{\nu,min}$ by more than 
$\pm 15\%$ of the latter value \label{plausible}}

\tablewidth{0pt} \tablehead{\colhead{Planet}&
\colhead{Plausible $\chi^2_\nu$ Ranges}&
\colhead{Plausible $T$ Ranges}}
\startdata
HD 37124 c & $28.62\leq\chi^2_\nu\leq 38.72$& $4.69\leq T\leq 7.12$ yr \\ 
HD 74156 c & $5.40\leq\chi^2_\nu\leq 7.31$&$5.85\leq T\leq 6.38$ yr \\
HD 38529 c & $1.17\leq\chi^2_\nu\leq 0.86$&$5.93\leq T\leq 6.08$ yr \\
55 Cnc d & $19.46\leq\chi^2_\nu\leq 26.33$&$12.99\leq T\leq 16.25$ yr \\
47 UMa c & $0.66\leq\chi^2_\nu\leq 0.90$&$5.91\leq T\leq 9.22$ yr \\
Jupiter (2 pc)& $1.07\leq\chi^2_\nu\leq 1.45$& $11.11\leq T\leq 11.61$ yr \\
Jupiter (3 pc)& $0.87\leq\chi^2_\nu\leq 1.18$& $10.98\leq T\leq 11.66$ yr \\
Jupiter (10 pc)& $0.70\leq\chi^2_\nu\leq 0.95$& $10.16\leq T\leq 12.07$ yr 
\enddata
\end{deluxetable}
\clearpage

\begin{deluxetable}{lccccc}

\tablecaption{Ratios of the rms errors $\sigma_m$ for a given
parameter ($M_p$, $a$, $T$, $e$, and $i$)
derived from  a multiple-planet fit (in simulations where the full
planetary system was generated around the parent star) to
the rms errors $\sigma_s$  derived from a single-planet solution
(in simulations where only one planet orbited the central star).
The values are averaged over the inclination angle \label{ratios}}

\tablewidth{0pt} \tablehead{\colhead{Planet}&
\colhead{$\sigma_m(M_p)/\sigma_s(M_p)$}&
\colhead{$\sigma_m(a)/\sigma_s(a)$}&
\colhead{$\sigma_m(T)/\sigma_s(T)$}&
\colhead{$\sigma_m(e)/\sigma_s(e)$}&
\colhead{$\sigma_m(i)/\sigma_s(i)$}}

\startdata
$\upsilon$ And b &2.42&2.32&1.56&0.98&1.36
\\ $\upsilon$ And c &1.42&1.35&1.59&1.40&1.35
\\ $\upsilon$ And d&1.35&1.37&1.54&1.52&1.35
\\ 55 Cnc b &1.39&1.39&1.49&1.35&1.23
\\ 55 Cnc c &1.48&1.48&1.40&1.14&1.18
\\ 55 Cnc d &1.00&0.76&0.83&1.53&1.71
\\ 47 UMa b &15.4&18.0&9.64&1.54&2.31
\\ 47 UMa c &1.04&1.09&1.12&3.69&2.71
\\ Gliese 876 b &1.21&1.21&1.21&1.39&1.12
\\ Gliese 876 c &1.22&1.23&1.55&1.14&1.12
\\ HD 12661 b &1.15&1.15&1.24&1.15&1.12
\\ HD 12661 c &1.19&1.21&1.37&1.33&1.10
\\ HD 160691 b &1.54&1.54&1.86&1.38&1.32
\\ HD 160691 c &2.09&2.07&2.00&1.78&1.66
\\ HD 168443 b &1.13&1.15&1.21&1.13&1.09
\\ HD 168443 c &0.46&0.47&0.86&0.97&1.09
\\ HD 37124 b &1.18&1.17&1.27&1.16&1.11
\\ HD 37124 c &0.28&0.48&1.00&0.84&0.86
\\ HD 38529 b &1.19&1.19&1.19&1.04&1.07
\\ HD 38529 c &0.86&0.85&0.78&0.55&0.87
\\ HD 74156 b &1.08&1.08&1.29&1.06&1.06
\\ HD 74156 c &0.59&0.71&0.94&0.68&0.66
\\ HD 82943 b &2.87&2.08&1.69&3.23&1.77
\\ HD 82943 c &2.50&1.81&1.95&2.05&1.70
\enddata
\end{deluxetable}
\clearpage

\begin{deluxetable}{cc}

\tablecaption{Results from a fit to the 
fractional errors on orbital elements and planet mass, 
assuming a three-dimensional power-law dependence on the 
three parameters $\alpha/\sigma_d$, $T$, and $e$ 
of the form: $\displaystyle 
\eta_0\times (\alpha/\sigma_d)^{\eta_1}\times 
T^{\eta_2}\times e^{\eta_3}$. The results are averaged 
over the inclination angle \label{powerlaw}}

\tablewidth{0pt} \tablehead{\colhead{Parameter Error (\%)}&
\colhead{Best-Fit Power-Law}}
\startdata
$\sigma_\mathrm{M_p}$ & 
$34.96\times (\alpha/\sigma_d)^{-0.72}\times T^{-0.42}\times 
e^{0.35}$\\
$\sigma_\mathrm{a}$ & 
$34.08\times (\alpha/\sigma_d)^{-0.73}\times T^{-0.43}\times 
e^{0.36}$\\
$\sigma_\mathrm{T}$ &
$2.50\times (\alpha/\sigma_d)^{-1.19}\times T^{2.65}\times 
e^{-0.12}$\\
$\sigma_\mathrm{e}$ & 
$23.39\times (\alpha/\sigma_d)^{-0.87}\times T^{0.01}\times 
e^{-1.06}$\\
$\sigma_\mathrm{i}$ & 
$24.28\times (\alpha/\sigma_d)^{-0.59}\times T^{-0.13}\times 
e^{0.08}$\\
$\sigma_\mathrm{\Omega}$ & 
$67.85\times (\alpha/\sigma_d)^{-0.43}\times T^{-0.06}\times 
e^{0.43}$
\enddata
\end{deluxetable}
\clearpage
\begin{figure}
\plotone{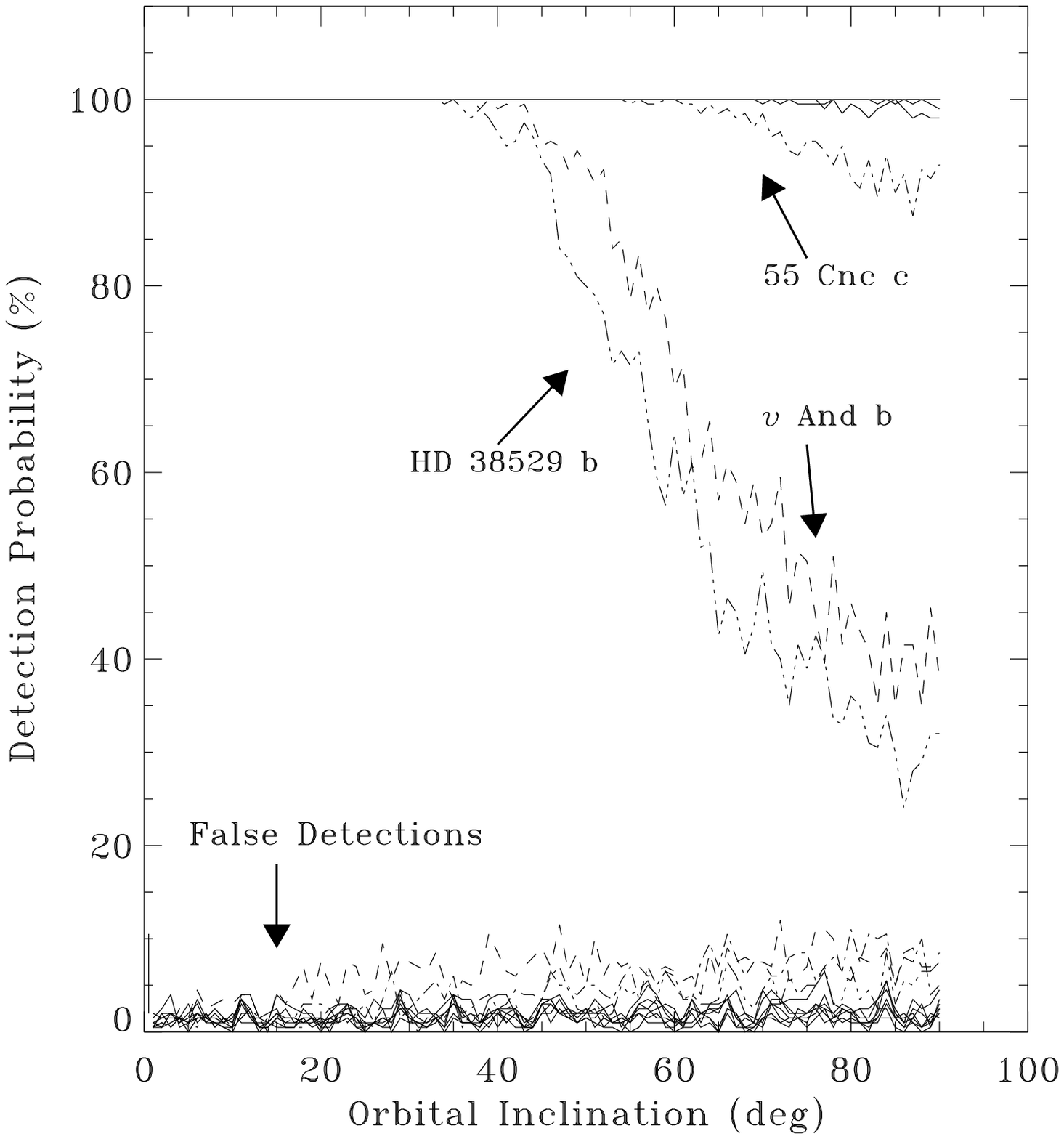}
\end{figure}
\clearpage
\begin{figure}
\plotone{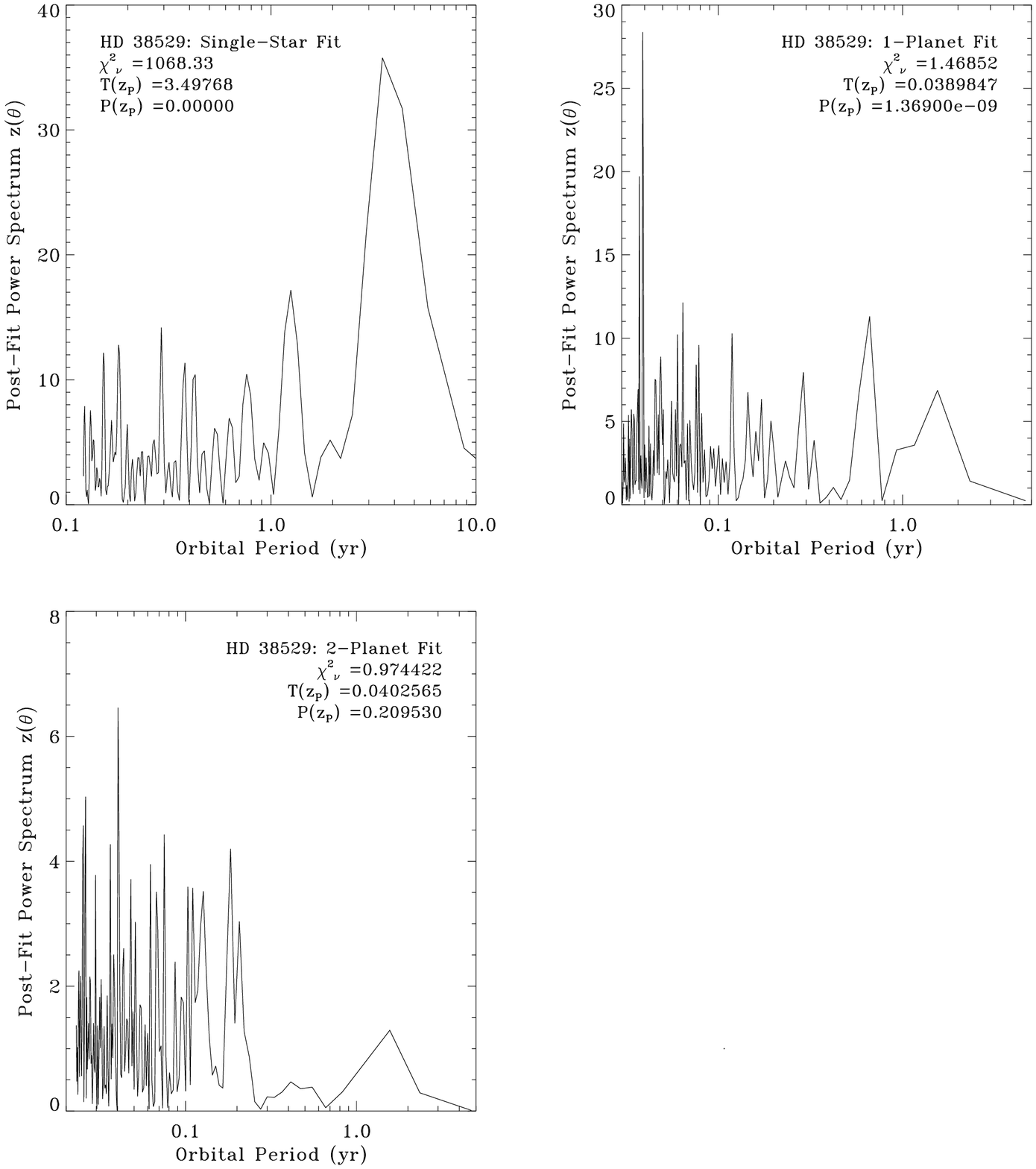}
\end{figure}
\clearpage
\begin{figure}
\plotone{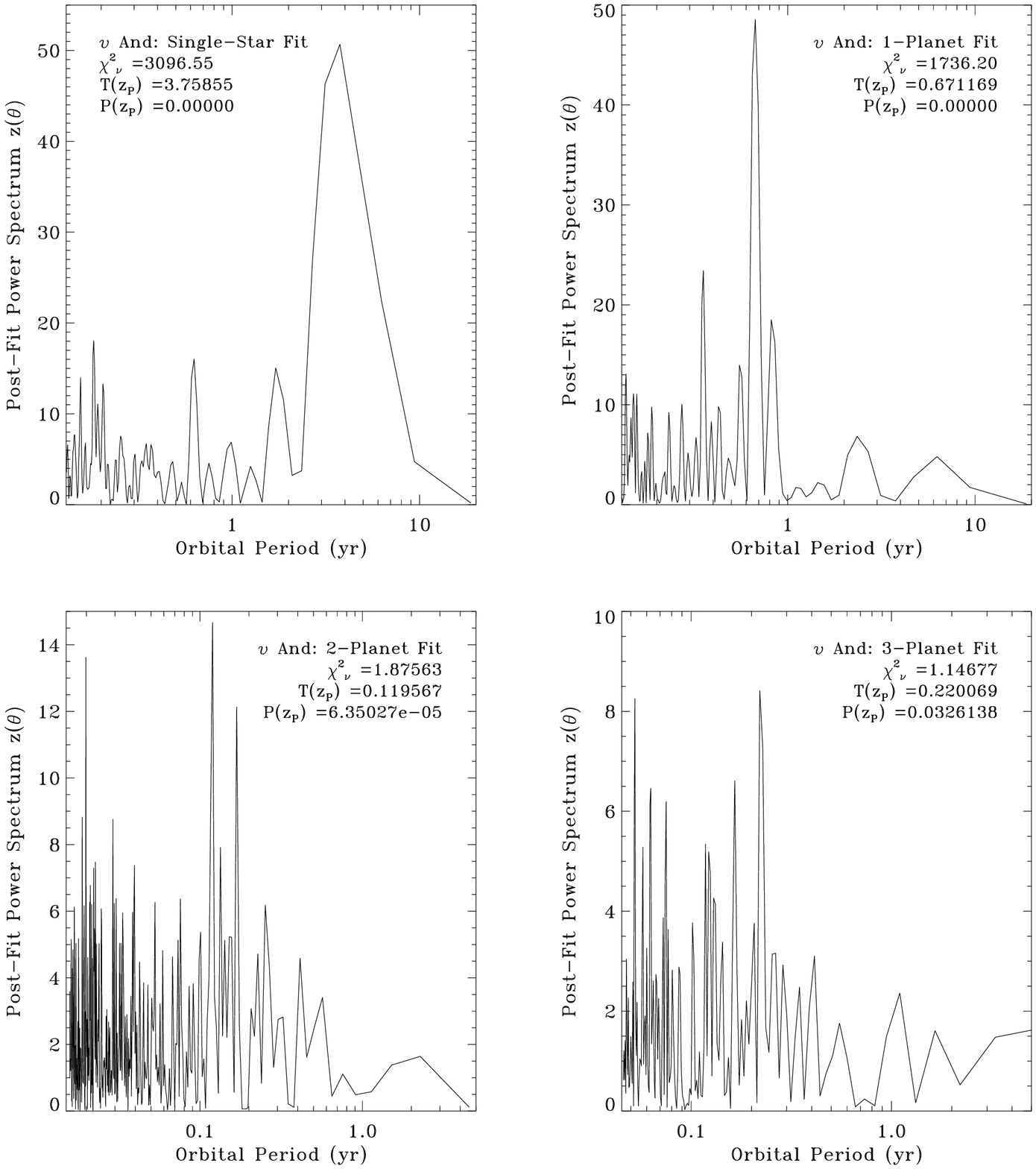}
\end{figure}
\clearpage
\begin{figure}
\plotone{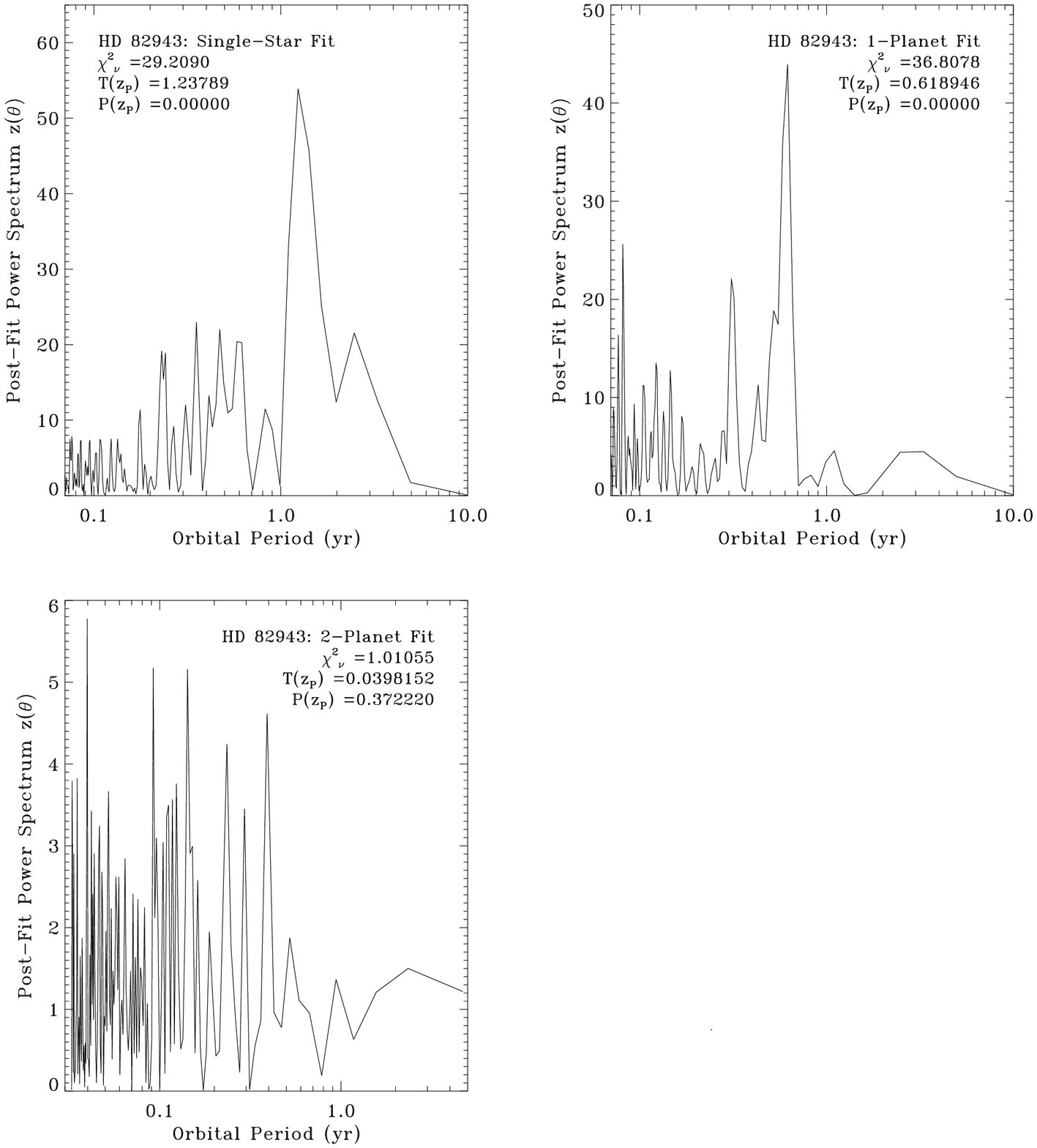}
\end{figure}
\clearpage
\begin{figure}
\plotone{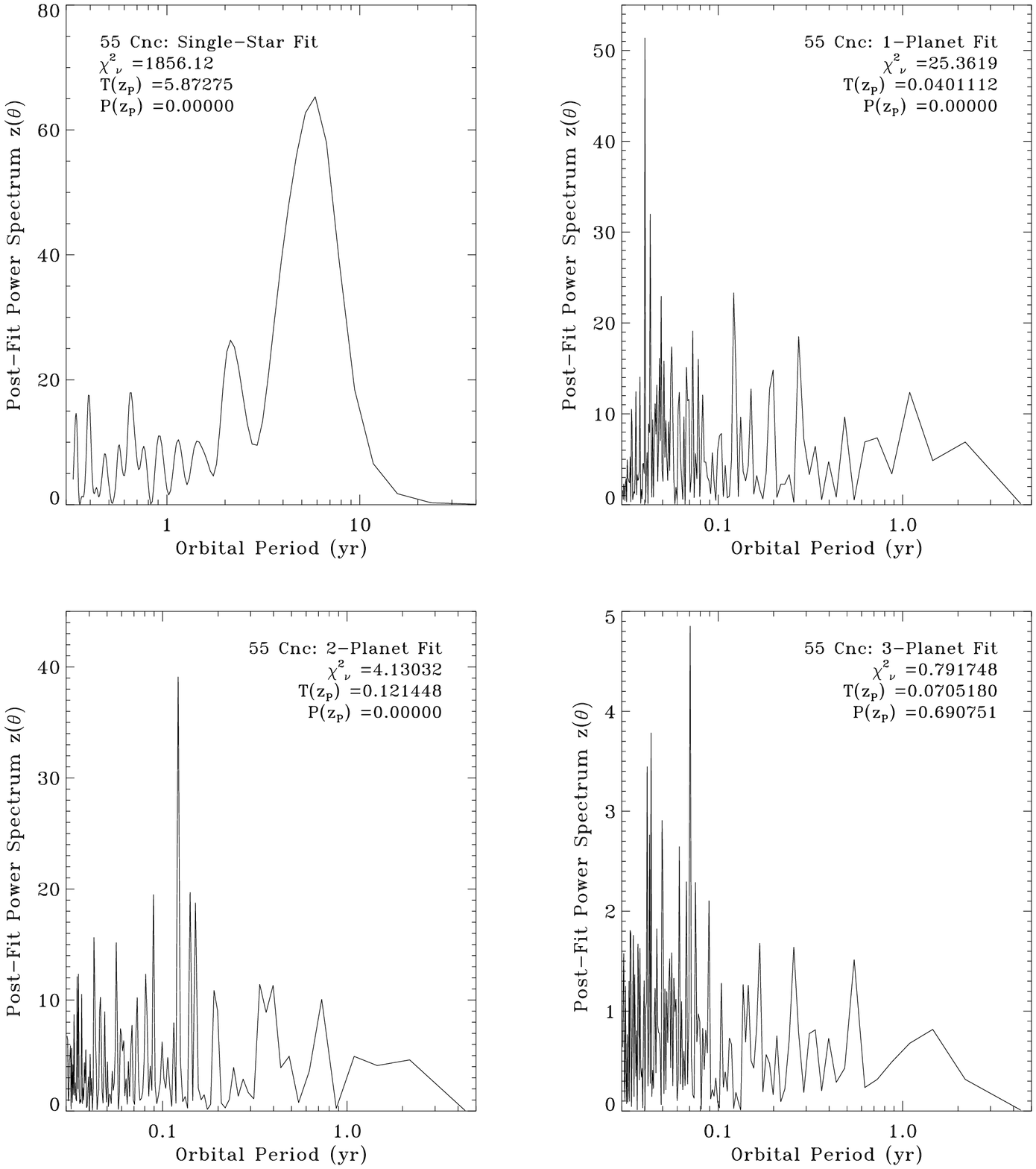}
\end{figure}
\clearpage
\begin{figure}
\plotone{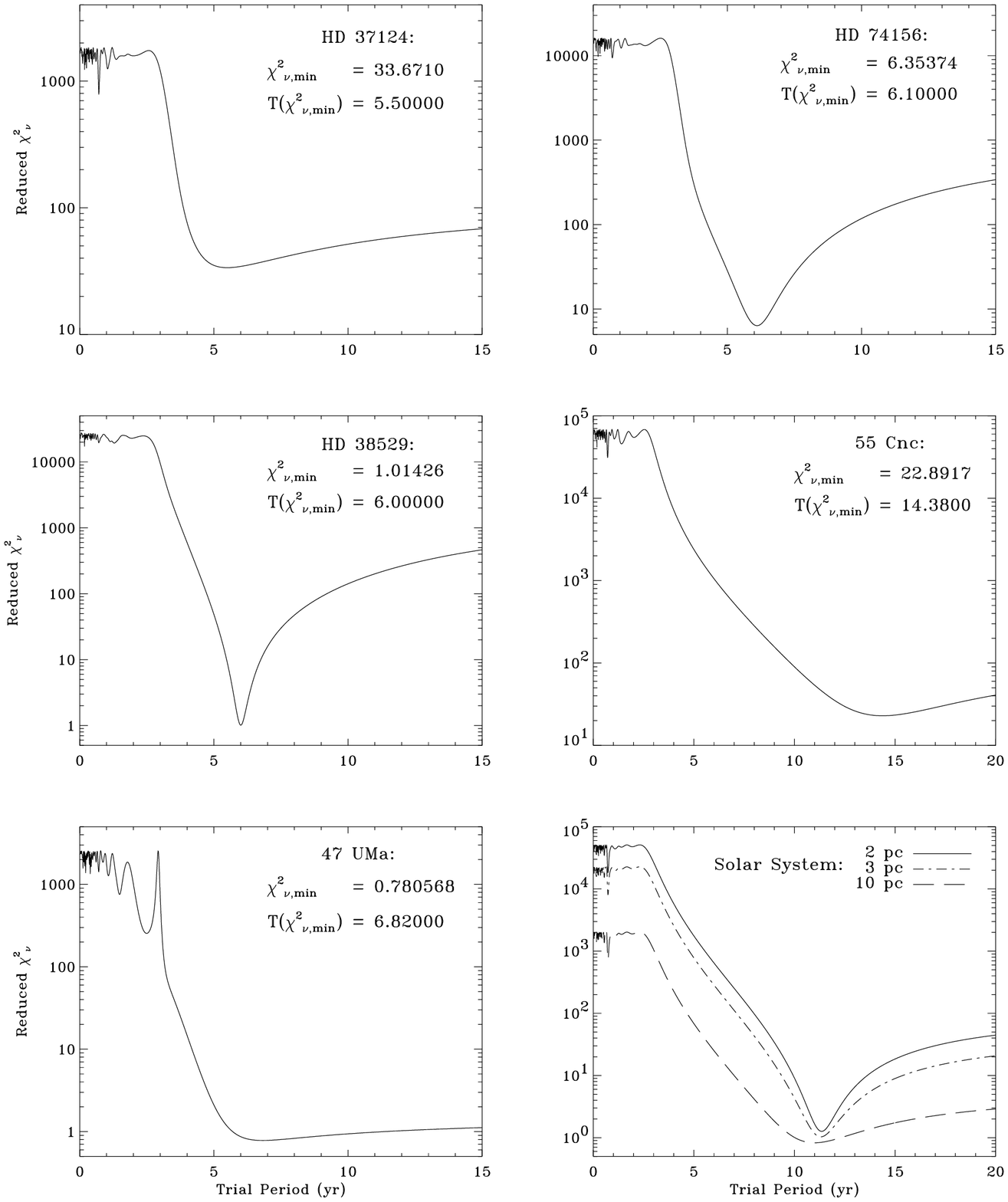}
\end{figure}
\clearpage
\begin{figure}
\plotone{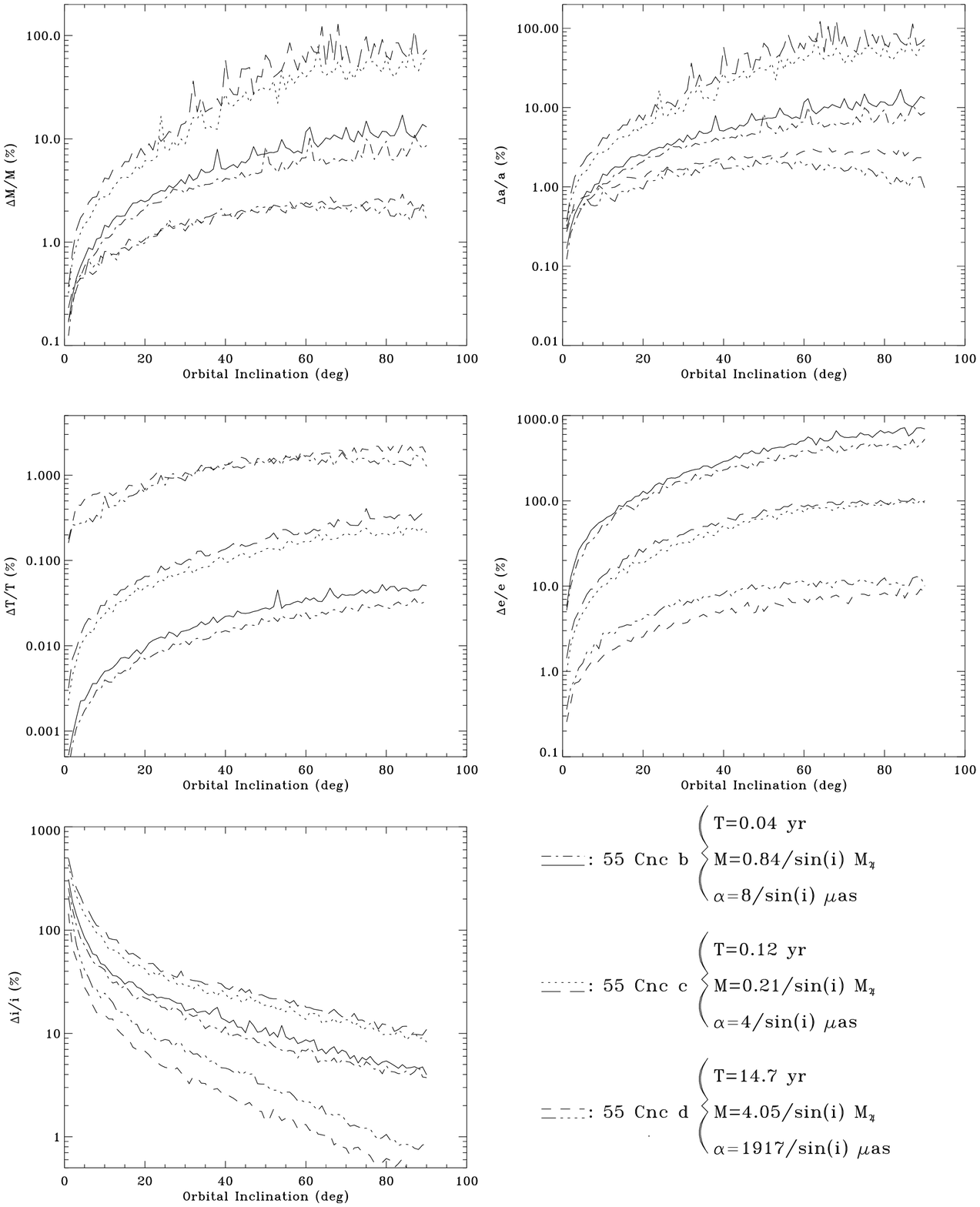}
\end{figure}
\clearpage
\begin{figure}
\plotone{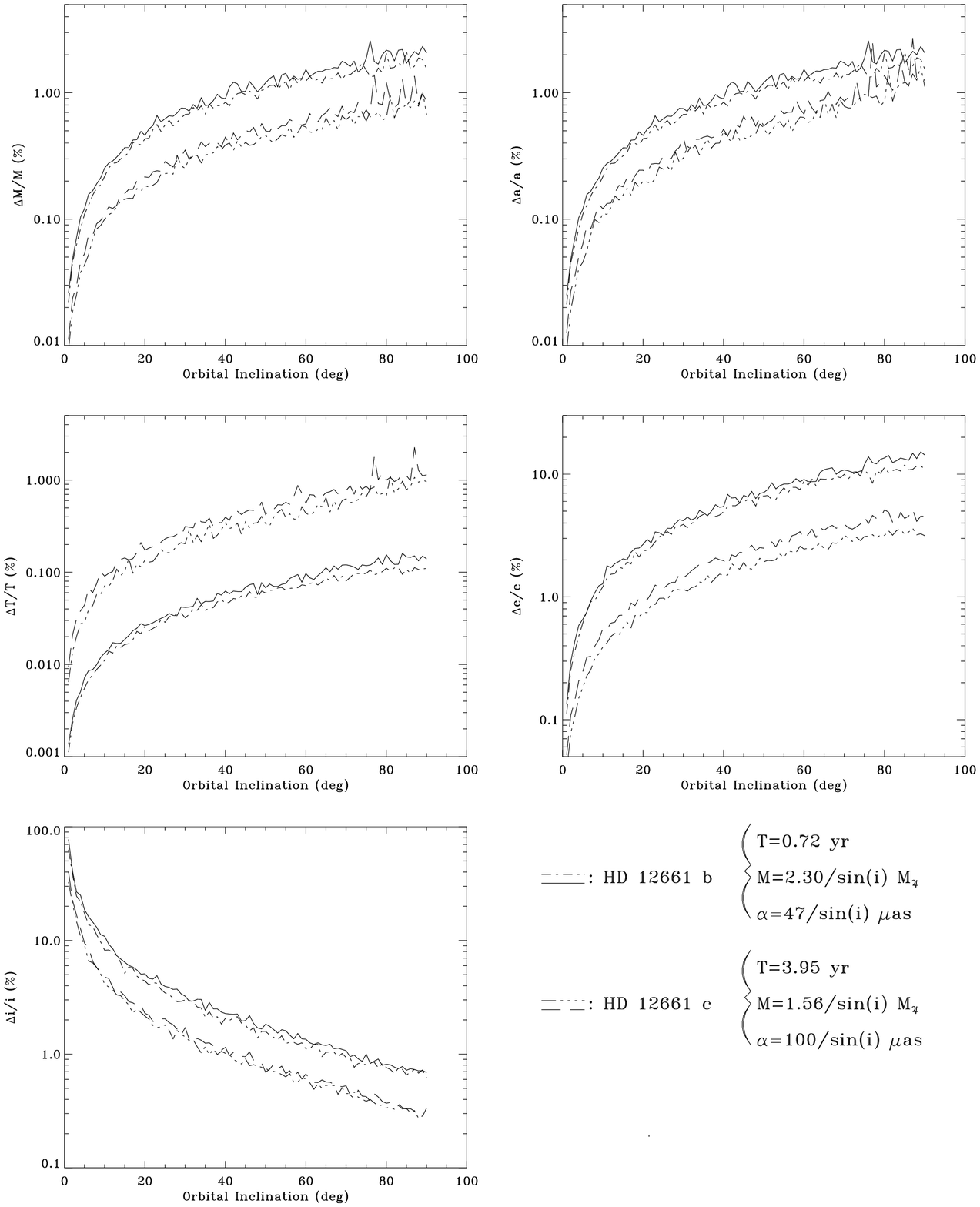}
\end{figure}
\begin{figure}
\plotone{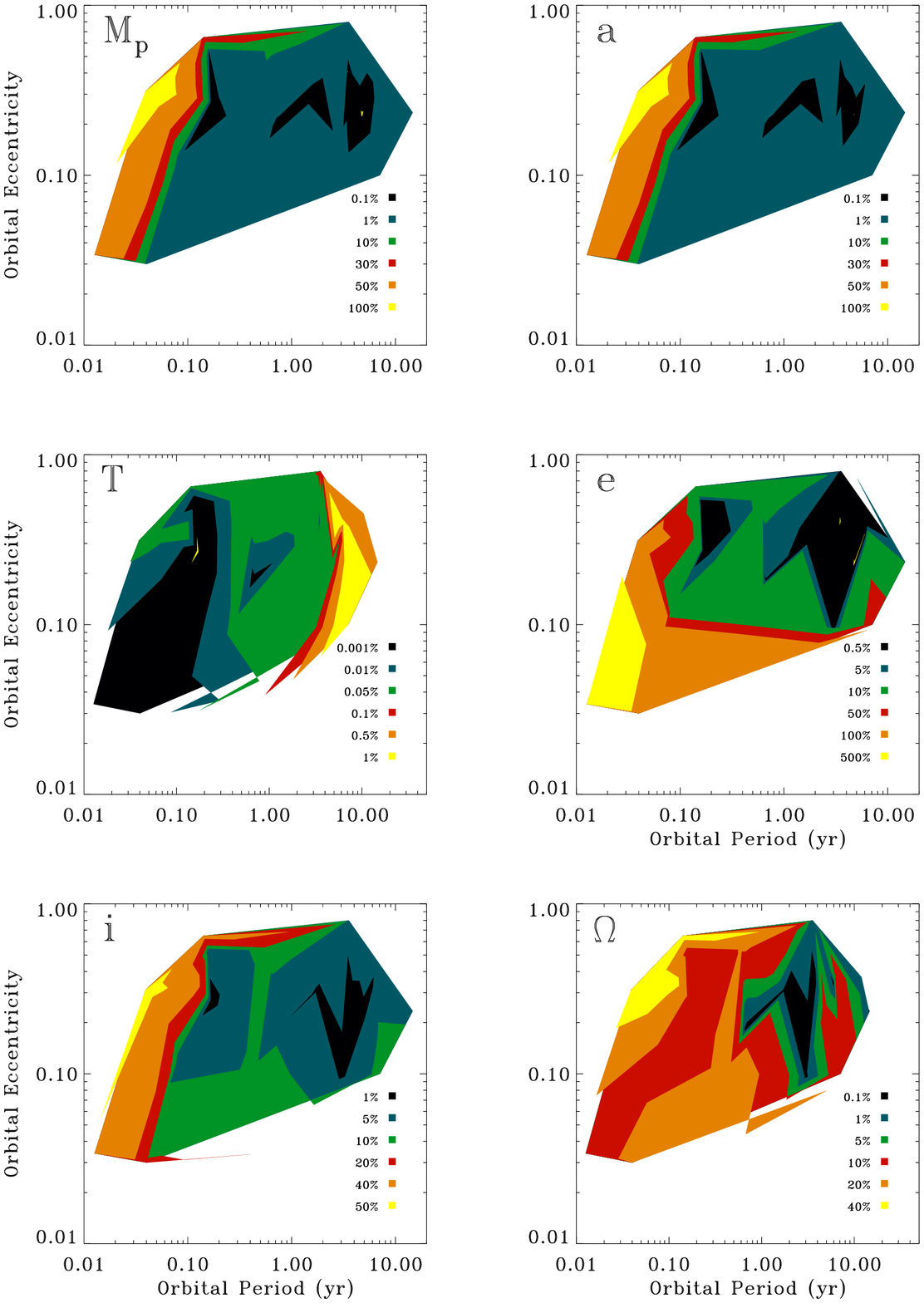}
\end{figure}
\clearpage
\begin{figure}
\plotone{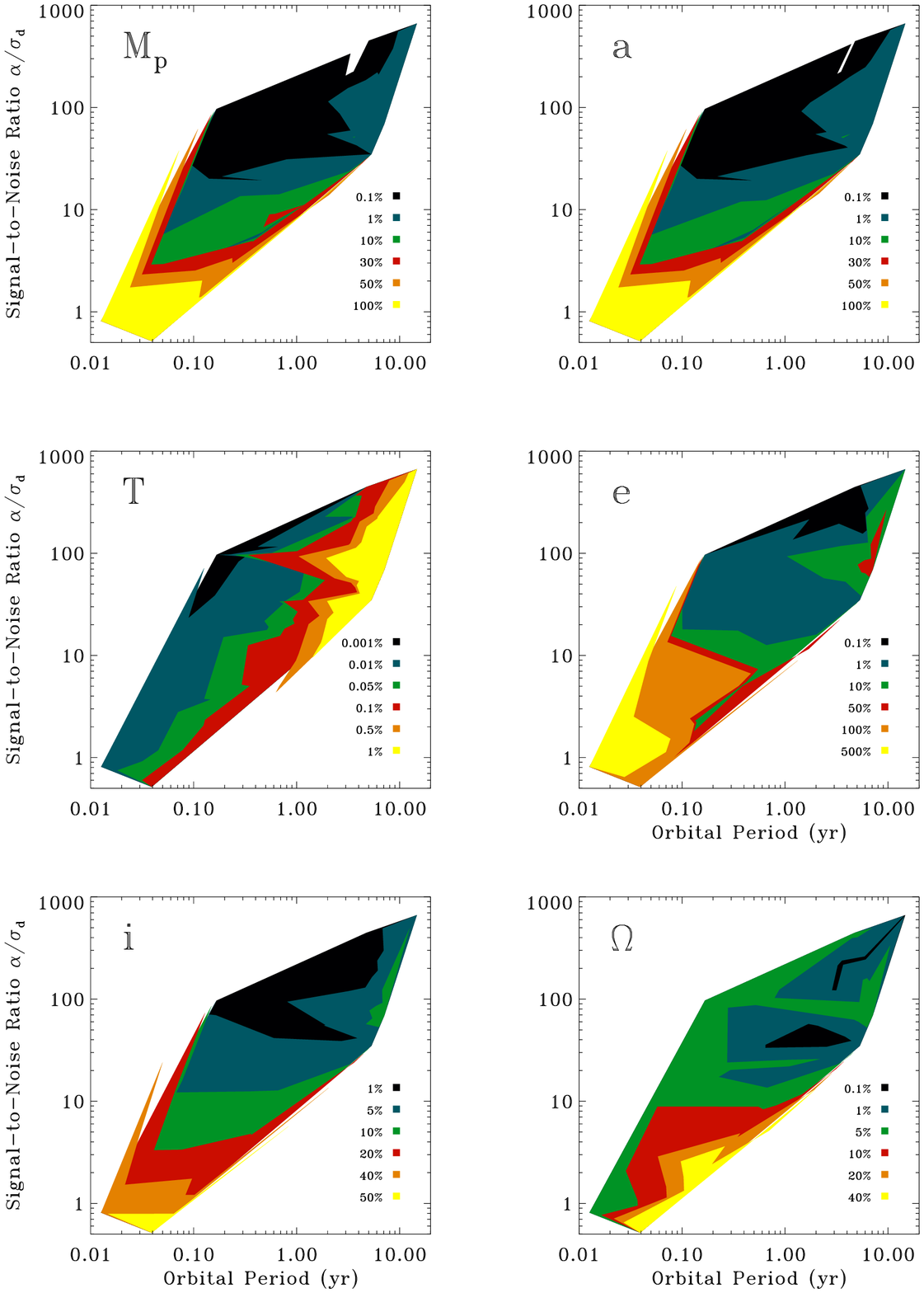}
\end{figure}
\clearpage
\clearpage
\begin{figure}
\plotone{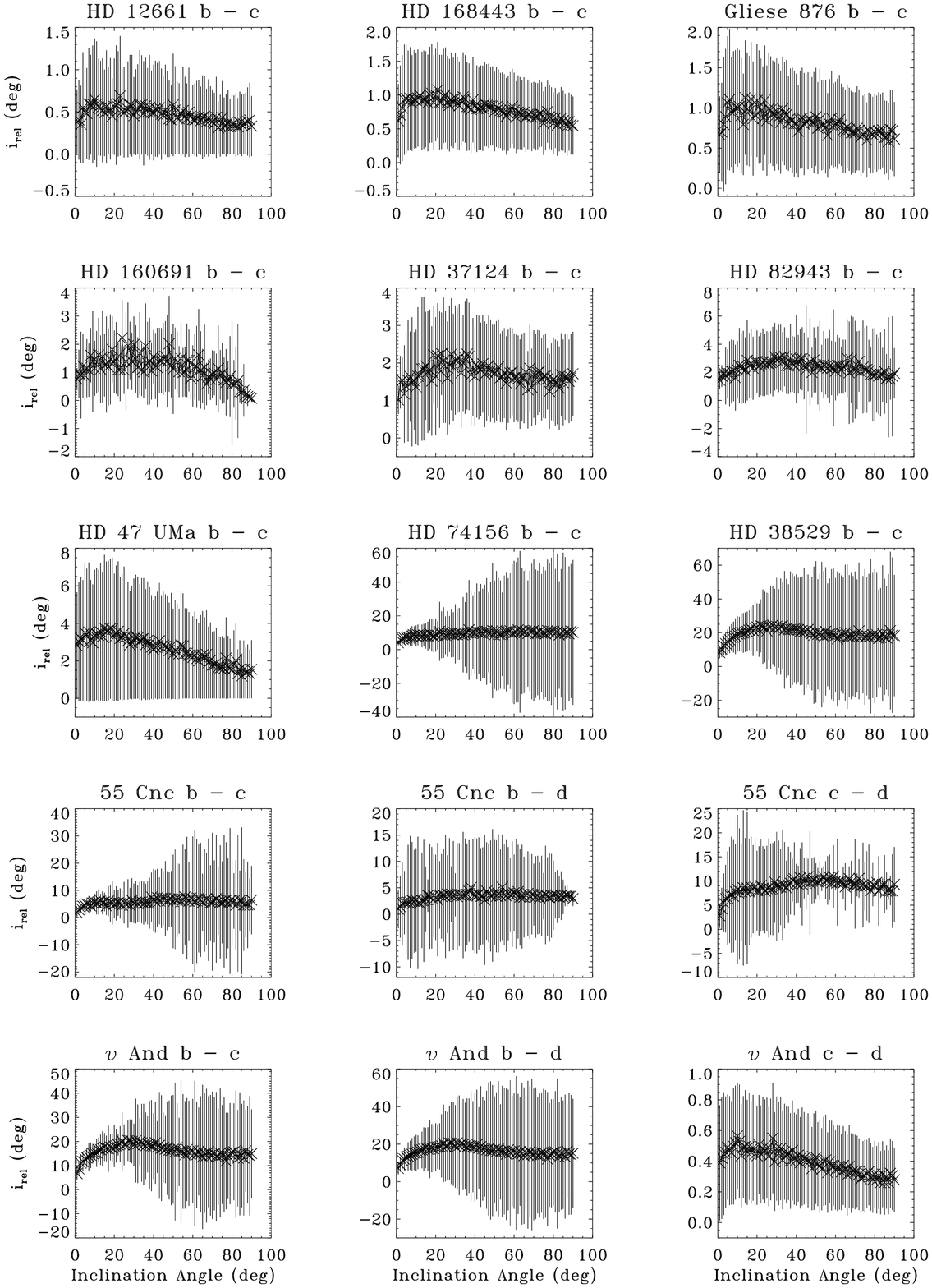}
\end{figure}
\clearpage
\begin{figure}
\epsscale{0.68}
\plotone{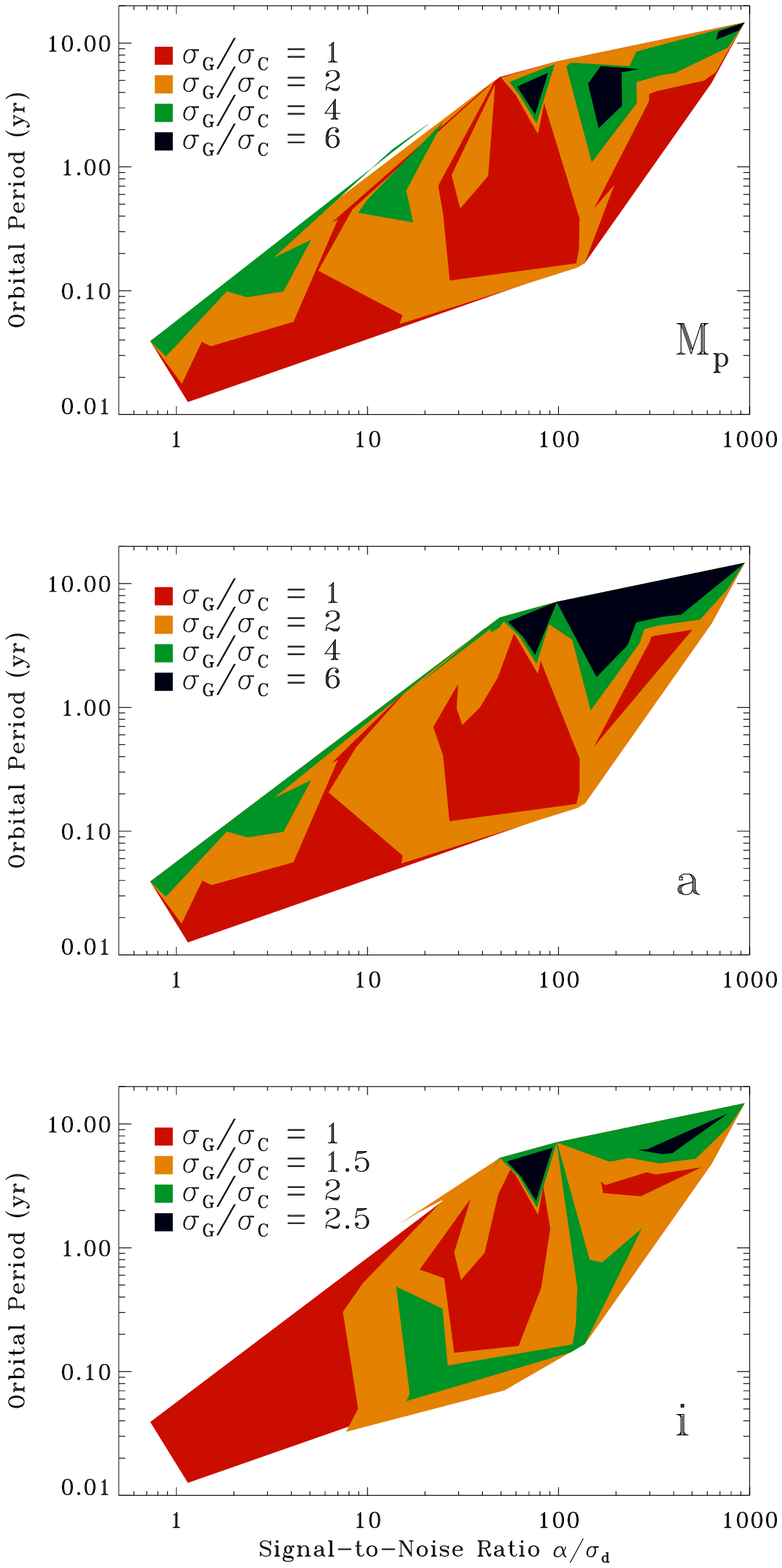}
\end{figure}
\clearpage
\begin{figure}
\epsscale{1.1}
\plottwo{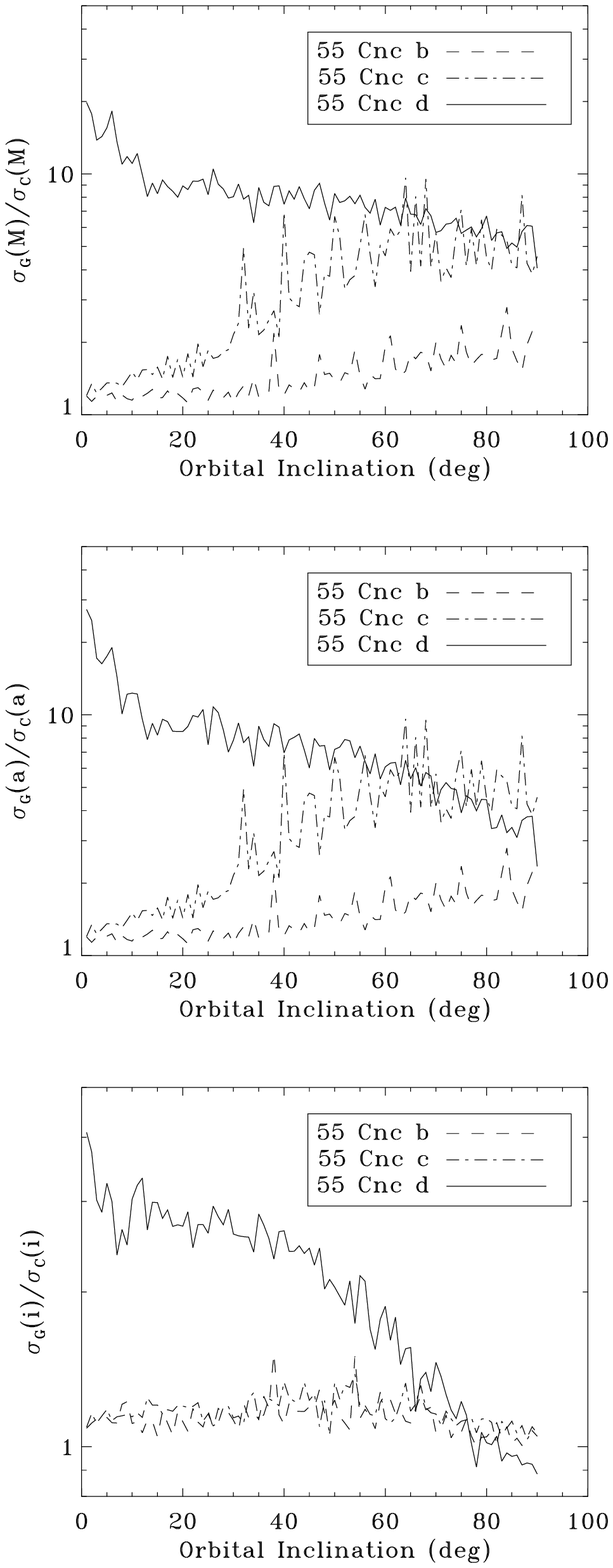}{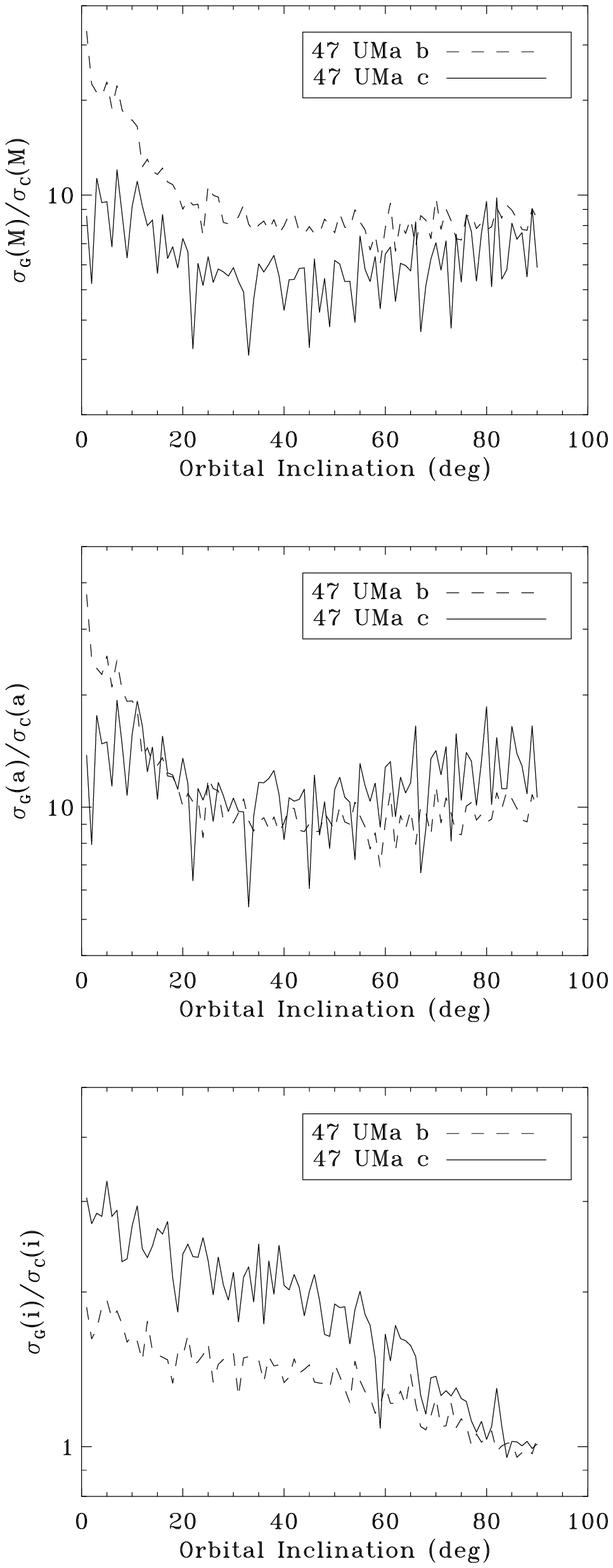}
\end{figure}
\clearpage
\begin{figure}
\epsscale{1.1}
\plottwo{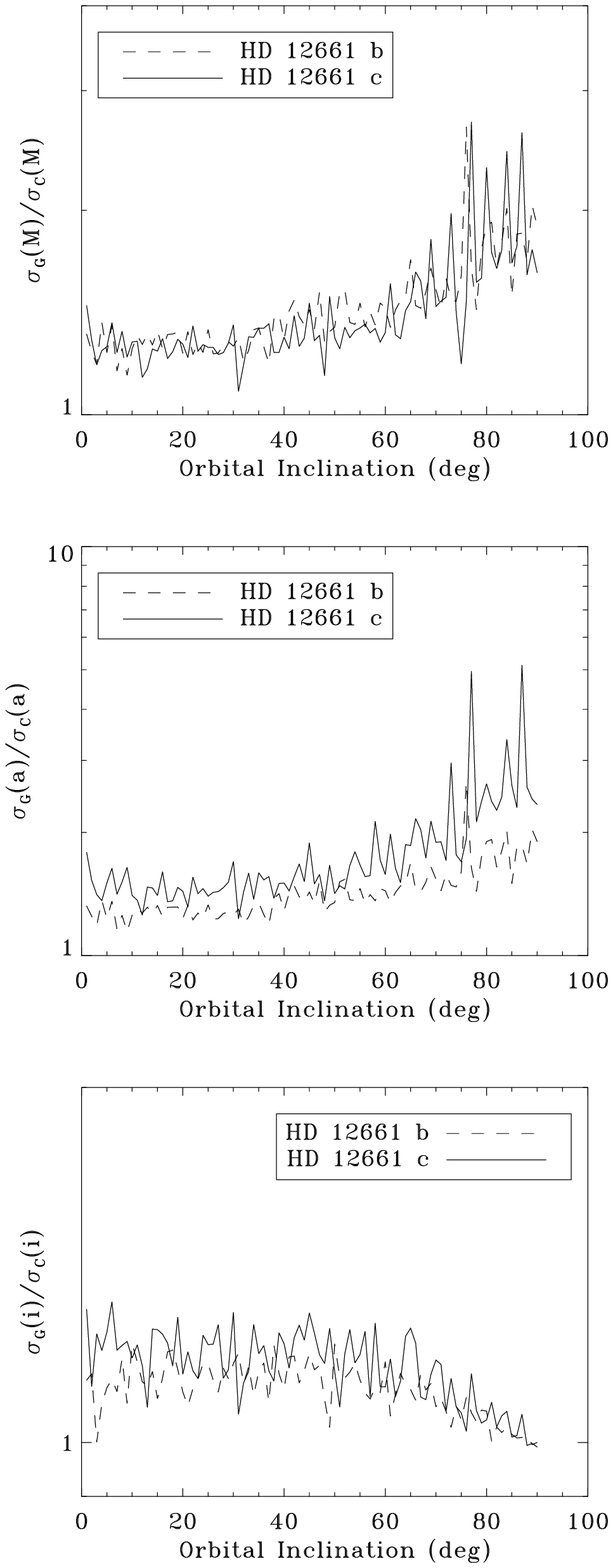}{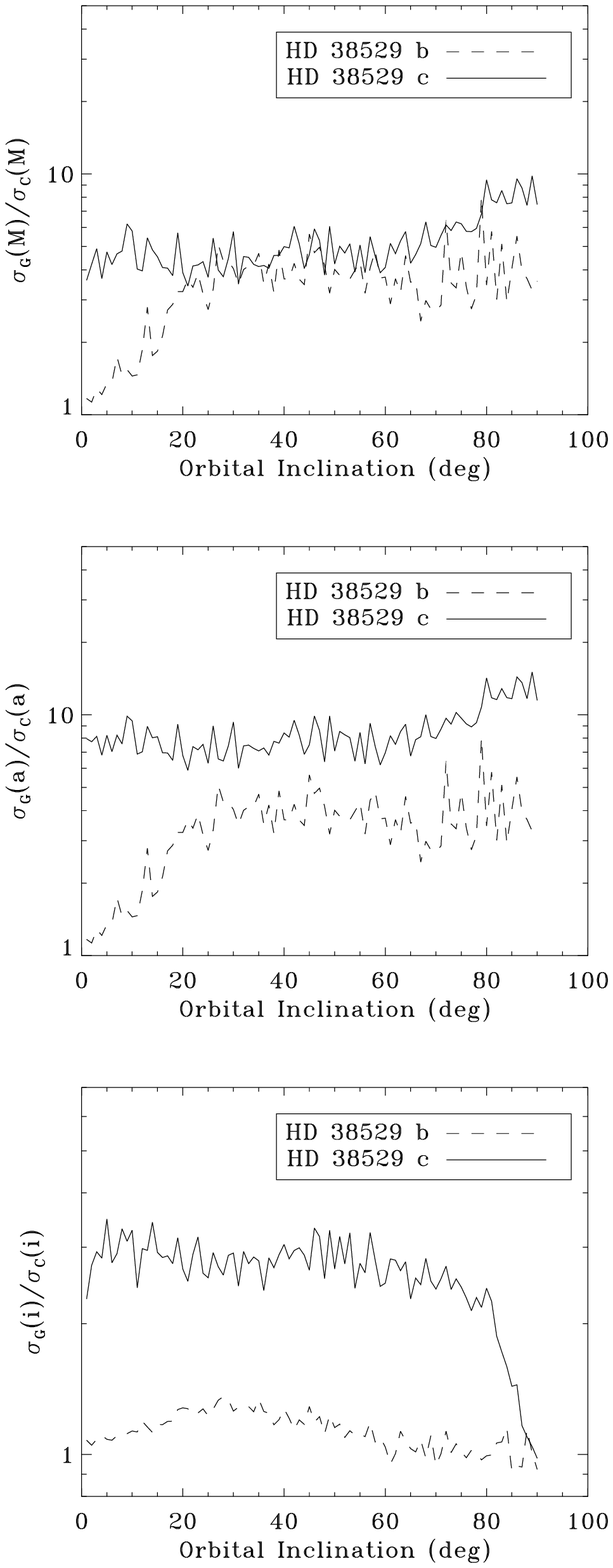}
\end{figure}
\clearpage





\newpage

\begin{thebibliography}{}
\bibitem[Artymowicz 2001]{arty01}
Artymowicz, P. 2001, \baas, 33, \#10.01
\bibitem[Bard 1974]{bard74}
Bard, Y. 1974, Nonlinear Parameter Estimation, New York: Academic Press 
\bibitem[Barnes \& Quinn 2001]{barnes01}
Barnes, R., Quinn, T. 2001, \apj, 550, 884
\bibitem[\protect\citeauthoryear{Benedict et al.}{2001}]{benedict01}
Benedict, G. F., et al. 2001, \aj, 121, 1607
\bibitem[Benedict et al. 2002]{benedict02}
Benedict, G. F., et al. 2002, \apjl, 581, L115
\bibitem[Bernstein 1997]{bernstein97}
Bernstein, H.-H. 1997, Proc. ESA Symp., 
Hipparcos - Venice '97, ESA SP-402, 705
\bibitem[Bevington \& Robinson 2003]{bevington03}
Bevington, P. R., Robinson, D. K. 2003, Data reduction and Error Analysis 
for the Physical Sciences - 3rd ed., Boston: McGraw-Hill
\bibitem[Black \& Scargle 1982]{black82}
Black, D. C., Scargle, J. D. 1982, \apj, 263, 854
\bibitem[Booth et al. 1999]{booth99}
Booth, A. J., et al. 1999, in Working on the Fringe: Optical and IR
Interferometry from Ground and Space, ed. S. Unwin \& R. Stachnik,
ASP Conf. Series, 194, 256
\bibitem[Boss 2000]{boss00}
Boss, A. P. 2000, \apjl, 536, L101
\bibitem[Boss 2001]{boss01}
Boss, A. P. 2001, \apj, 563, 367
\bibitem[Butler et al. 1999]{butler99}
Butler, R. P., et al. 1999, \apj, 526, 916
\bibitem[Butler et al. 2003]{butler03}
Butler, R. P., Marcy, G. W., Vogt, S. S., Fischer, D. A.,
Henry, G. W., Laughlin G., Wright, J. 2003, \apj, 582, 455
\bibitem[Casertano et al. 1996]{caser96}
Casertano, S., Lattanzi, M. G., Perryman, M. A. C., Spagna, A. 1996, 
\apss, 241, 89
\bibitem[Casertano \& Sozzetti 1999]{caser99}
Casertano, S., Sozzetti, A. 1999, in Working on the Fringe: Optical and IR
Interferometry from Ground and Space, ed. S. Unwin \& R. Stachnik,
ASP Conf. Series 194, 171
\bibitem[Charbonneau et al. 2000]{charbon00}
Charbonneau, D., Brown, T. M., Latham, D. W., Mayor, M. 2000, \apjl,
529, L45
\bibitem[Charbonneau et al. 2002]{charbon02}
Charbonneau, D., Brown, T. M., Noyes, R. W., Gilliland, R. L. 2002, \apj,
568, 377
\bibitem[Chiang et al. 2001]{chiang01}
Chiang, E. I., Tabachnik, S., Tremaine, S. 2001, \aj, 122, 1607
\bibitem[Chiang \& Murray 2002]{chiang02}
Chiang, E. I., Murray, N. 2002, \apj, 576, 473
\bibitem[Colavita et al. 1999]{colavita99}
Colavita, M. M., et al. 1999, \apj, 510, 505
\bibitem[\protect\citeauthoryear{Cumming et al.}{1999}]{cumming99}
Cumming, A., Marcy, G. W., Butler, R. P. 1999, \apj, 526, 890
\bibitem[Cuntz et al. 2003]{cuntz03}
Cuntz, M., Von Bloh, W., Bounama, C., Franck, S. 2003, Icarus, 162, 215
\bibitem[Danner \& Unwin 1999]{danner99}
Danner, R., Unwin, S. 1999, SIM: Taking the Measure of the Universe, NASA/JPL
\bibitem[\protect\citeauthoryear{Eisner \& Kulkarni}{2001a}]{eisner01a}
Eisner, J. A., Kulkarni, S. R. 2001a, \apj, 550, 871
\bibitem[\protect\citeauthoryear{Eisner \& Kulkarni}{2001b}]{eisner01b}
Eisner, J. A., Kulkarni, S. R. 2001b, \apj, 561, 1107
\bibitem[\protect\citeauthoryear{Eisner \& Kulkarni}{2002}]{eisner02}
Eisner, J. A., Kulkarni, S. R. 2002, \apj, 574, 426
\bibitem[Fischer et al. 2001]{fischer01}
Fischer, D. A., Marcy, G W., Butler, R. P., Vogt, S. S.,
Frink, S., Apps, K. 2001, \apj, 551, 1107
\bibitem[Fischer et al. 2002]{fischer02}
Fischer, D. A., Marcy, G. W., Butler, R. P., Laughlin, G., Vogt, S. S.
2002, \apj, 564, 1028
\bibitem[Fischer et al. 2003]{fischer03}
Fischer, D. A., et al. 2003, \apj, 586, 1394
\bibitem[\protect\citeauthoryear{Franz et al.}{1998}]{franz98}
Franz, O. G., et al. 1998, \aj, 116, 1432
\bibitem[Gilliland \& Baliunas 1987]{gilli87}
Gilliland, R. L., Baliunas, S. L. 1987, \apj, 314, 766
\bibitem[Gladman 1993]{gladman93}
Gladman, B. 1993, Icarus, 106, 247
\bibitem[God\'zdziewski \& Maciejewski 2001]{godz01}
God\'zdziewski, K., Maciejewski, A. J. 2001, \apj, 563, L81
\bibitem[God\'zdziewski et al. 2002]{godzetal02}
God\'zdziewski, K., Bois, E., Maciejewski, A. J. 2002, \mnras, 332, 839
\bibitem[God\'zdziewski 2002]{godz02}
God\'zdziewski, K. 2002, \aap, 393, 997
\bibitem[God\'zdziewski 2003a]{godz03a}
God\'zdziewski, K. 2003a, \aap, 398, 315
\bibitem[God\'zdziewski 2003b]{godz03b}
God\'zdziewski, K. 2003b, \aap, 398, 1151
\bibitem[God\'zdziewski \& Maciejewski 2003]{godzmac03}
God\'zdziewski, K., Maciejewski, A. J. 2003, \apj, 586, L153
\bibitem[Green 1985]{green85}
Green, R. 1985, Spherical Astronomy, Cambridge University Press
\bibitem[\protect\citeauthoryear{Guirado et al.}{1997}]{guirado97}
Guirado, J. C., et al. 1997, \apj, 490, 835
\bibitem[Heintz 1978]{heintz78}
Heintz, W. D. 1978, Double Stars, Boston D. Reidel Publishing Company
\bibitem[Henry et al. 2000]{henry00}
Henry, G. W., Marcy, G. W., Butler, R. P., Vogt, S. S. 2000,
\apjl, 529, L41
\bibitem[\protect\citeauthoryear{Horne \& Baliunas}{1986}]{horne86}
Horne, J. H., Baliunas, S. L. 1986, \apj, 302, 757
\bibitem[Jensen \& Ulrych 1973]{jensen73}
Jensen, O. G., Ulrych, T. 1973, \aj, 78, 1104
\bibitem[Ji et al. 2002a]{ji02}
Ji, J., Li., G., Liu, L. 2002, \apj, 572, 1041
\bibitem[Ji et al. 2003]{ji03}
Ji, J., Kinoshita, H., Liu, L., Li, G. 2003, \apj, 585, L139
\bibitem[Jiang \& Ip 2001]{jiang01}
Jiang, I., Ip, W. 2001, \aap, 367, 943
\bibitem[Jones et al. 2001]{sleep01}
Jones, B. W., Sleep, P. N., Chambers, J. E. 2001, \aap, 366, 254
\bibitem[Jones \& Sleep 2002a]{sleep02a}
Jones, B. W., Sleep, P. N. 2002a, \aap, 393, 1015
\bibitem[Jones \& Sleep 2002b]{sleep02b}
Jones, B. W., Sleep, P. N. 2002b, in Scientific Frontiers
in Research on Extrasolar Planets, ed. D. Deming and S. Seager, 
ASP Conf. Series, 294, 225
\bibitem[Jones et al. 2002]{jones02}
Jones, H. R. A., Butler, R. P., Marcy, G. W., Tinney, C. G., Penny, A. J.,
McCarthy, C., Carter, B. D. 2002, \mnras, 337, 1170
\bibitem[Kasting et al. 1993]{kasting93}
Kasting, J. F., Whitmire, D. P., Reynolds, R. T. 1993,
Icarus, 101, 108
\bibitem[Kells et al. 1942]{kells42}
Kells, L. M., Kern, W. F., Bland, J. R. 1942, Spherical Trigonometry 
with Naval and Military Applications, New York: McGraw-Hill
\bibitem[Kinoshita \& Nakai 2001]{kino01}
Kinoshita, H., Nakai, H. 2001, \pasj, 53, L25
\bibitem[Kiseleva-Eggleton et al. 2002]{kiseleva02}
Kiseleva-Eggleton, L., Bois, E., Rambaux, N., Dvorak, R. 2002,
\apj, 578, L145
\bibitem[Konacki et al. 2002]{konacki02}
Konacki, M., Maciejewski, A. J., Wolszczan, A. 2002, \apj, 567, 566
\bibitem[Konacki et al. 2003]{konacki03}
Konacki, M., Torres, G., Jha, S., Sasselov, D. 2003, \nat, 421, 507
\bibitem[Laskar 1994]{laskar94}
Laskar, J. 1994, \aap, 287, L9
\bibitem[Lattanzi et al. 2000]{lattanzi00}
Lattanzi, M. G., Spagna, A., Sozzetti, A., Casertano, S. 2000,
\mnras, 317, 211
\bibitem[Laughlin \& Adams 1999]{laugh99}
Laughlin, G., Adams, F. C. 1999, \apj, 526, 881
\bibitem[Laughlin \& Chambers 2001]{laugh01}
Laughlin, G., Chambers, J. E. 2001, \apj, 551, L109
\bibitem[Laughlin et al. 2002]{laugh02}
Laughlin, G., Chambers, J. E., Fischer, D. A. 2002, \apj, 579, 455
\bibitem[Lee \& Peale 2002a]{lee02a}
Lee, M. H., Peale, S. J. 2002a, \apj, 567, 596
\bibitem[Lee \& Peale 2002b]{lee02b}
Lee, M. H., Peale, S. J. 2002b, in Scientific Frontiers
in Research on Extrasolar Planets, ed. D. Deming and S. Seager, 
ASP Conf. Series, 294, 197
\bibitem[Lissauer 1993]{lissauer93}
Lissauer, J. J. 1993, ARA\&A, 31, 129
\bibitem[1989]{marcy89}
Marcy, G. W., Benitz, K. J. 1989, \apj, 344, 441
\bibitem[Marcy et al. 2001a]{marcy01a} Marcy G. W., Butler R. P., Fischer D.,
      et al. 2001a, \apj, 556, 296
\bibitem[Marcy et al. 2001b]{marcy01b} Marcy G. W., Butler R. P., Vogt S. S.,
et al. 2001b, \apj, 555, 418
\bibitem[Marcy et al. 2002a]{marcy02a}
Marcy, G. W., Butler, R. P., Fischer, D. A., Vogt, S. S. 2002a, in
Scientific Frontiers in Research on Extrasolar Planets,
ed. D. Deming and S. Seager, ASP Conf. Series, 294, 1
\bibitem[Marcy et al. 2002b]{marcy02b}
Marcy, G. W., Butler, R. P., Fischer, D. A., Laughlin, G., Vogt, S. S.,
Henry, G. W., Pourbaix, D. 2002b, \apj, 581, 1375 
 \bibitem[Mariotti et al. 1998]{mariotti98}
Mariotti, J. M., et al. 1998, \procspie\, 3350,
Astronomical Interferometry, Ed. R. D. Reasenberg, p. 800
\bibitem[\protect\citeauthoryear{Mayer et al.}{2002}]{mayer02}
Mayer, L., Quinn, T., Wadsley, J., Stadel, J. 2002, Science, 298, 1756
\bibitem[Mayor et al. 2001]{mayor01}
Mayor, M., et al. 2001, ESO Press Release 07/01
\bibitem[Mazeh et al. 2000]{mazeh00}
Mazeh, T., et al. 2000, \apjl, 532, L55
\bibitem[Menou \& Tabachnik 2003]{menou03}
Menou, C., Tabachnik, S. 2003, \apj, 583, 473 
\bibitem[Monet 1983]{monet83}
Monet, D. G. 1983, in Current Techniques in Double and 
Multiple Star Research, ed. R. S. Harrington and O. G. Franz, 
IAU Coll., 62, 286 
\bibitem[Nagasawa et al. 2003]{nagasawa03}
Nagasawa, M., Lin, D. N. C., Ida, S. 2003, \apj, 586, 1374
\bibitem[Nelson \& Angel 1998]{nelson98}
Nelson, A. F., Angel, J. R. P. 1998, \apj, 500, 940
\bibitem[Nelson \& Papaloizou 2002]{nelson02}
Nelson, R. P., Papaloizou, J. C. B. 2002, \mnras, 333, L26
\bibitem[Noble et al. 2002]{noble02}
Noble, M., Musielak, Z. E., Cuntz, M. 2002, \apj, 572, 1024
\bibitem[Novak et al. 2002]{novak02}
Novak, G. S., Lai, D., Lin, D. N. C. 2002, in Scientific Frontiers
in Research on Extrasolar Planets, ed. D. Deming and S. Seager, 
ASP Conf. Series, 294, 177
\bibitem[Perryman et al. 2001]{perryman01}
Perryman, M. A. C., et al. 2001, \aap, 369, 339
\bibitem[Pollack et al. 1996]{pollack96}
Pollack, J. B., Hubickyj, O., Bodenheimer, P., Lissauer, J. J.,
Podolack, M., Greenzweig, Y. 1996, Icarus, 124, 62
\bibitem[Pourbaix \& Jorissen 2000]{pourbaix00}
Pourbaix, D., Jorissen, A. 2000, \aaps, 145, 161
\bibitem[1989]{press89}
Press, W. H., Rybycki, G. B. 1989, \apj, 338, 277
\bibitem[Press et al. 1992]{press92}
Press, W. H., Teukolsky, S. A., Vetterling, V. T., Flannery, B. P. 1992,
Numerical Recipes in FORTRAN: the Art of Scientific Computing,
Cambridge University Press
\bibitem[Queloz et al. 2000]{queloz00}
Queloz, D., et al. 2000, \aap, 359, L13
\bibitem[Rivera \& Lissauer 2000]{rivera00}
Rivera, E. J., Lissauer, J. J. 2000, \apj, 530, 454
\bibitem[Rivera \& Lissauer 2001a]{rivera01a}
Rivera, E. J., Lissauer, J. J. 2001a, \apj, 554, 1141
\bibitem[Rivera \& Lissauer 2001b]{rivera01b}
Rivera, E. J., Lissauer, J. J. 2001b, \apj, 558, 392
\bibitem[Rivera \& Haghighipour 2002]{rivera02}
Rivera, E. J., Haghighipour N. 2002, in Scientific Frontiers
in Research on Extrasolar Planets, ed. D. Deming and S. Seager, 
ASP Conf. Series, 294, 205
\bibitem[Scargle 1982]{scargle82}
Scargle, J. D. 1982, \apj, 263, 835
\bibitem[Schwarzenberg-Czerny 1996]{czerny96}
Schwarzenberg-Czerny, A. 1996, \apjl, 460, L107
\bibitem[Snellgrove et al. 2001]{snell01}
Snellgrove, M. D., Papaloizou, J. C. B., Nelson, R. P. 2001, \aap, 374, 1092
\bibitem[\protect\citeauthoryear{Sozzetti et al.}{2001}]{sozzetti01}
Sozzetti, A., Casertano, S., Lattanzi, M. G., Spagna A. 2001, \aap, 373, L21
\bibitem[Sozzetti et al. 2002]{sozzetti02}
Sozzetti, A., Casertano, S., Brown, R. A., Lattanzi, M. G. 2002, \pasp,
114, 1173 (S02)
\bibitem[Stepinski et al. 2000]{stepinski00}
Stepinski, T. F., Malhotra, R., Black, D.C. 2000, \apj, 545, 1044
\bibitem[Sussman \& Wisdom 1992]{sussman92}
Sussman, G. J., Wisdom, J. 1992, Science, 257, 56
\bibitem[Th\'ebault et al. 2002]{thebault02}
Th\'ebault, P., Marzari, F., Scholl, H. 2002, \aap, 384, 594
\bibitem[Udry et al. 2002]{udry02}
Udry, S., Mayor, M., Naef, D., Pepe, F., Queloz, D.,
Santos, N. C., Burnet, M. 2002, \aap, 390, 267
\bibitem[Walker et al. 1995]{walker95}
Walker, G. A. H., Walker, A. R., Irwin, A. W., Larson, A. M.,
Yang, S. L. S., Richardson, D. C. 1995, Icarus, 116, 359
\bibitem[1992]{wols92}
Wolszczan, A., Frail, D. A. 1992, \nat, 355, 145
\end{thebibliography}
\end{document}